\documentclass[10pt, a4paper]{article}
\usepackage[english]{babel}
\usepackage[latin1]{inputenc}
\usepackage[tbtags]{amsmath}
\usepackage{amsfonts,euscript,amssymb,stmaryrd,braket}
\usepackage{graphics,tikz}
\usetikzlibrary{arrows,decorations.markings,patterns}
\usepackage{slashed}
\definecolor{hyperref}{RGB}{026,028,185}
\usepackage[bookmarks=true,colorlinks=true,linkcolor=hyperref,citecolor=hyperref,urlcolor=hyperref,bookmarksnumbered]{hyperref}

\numberwithin{equation}{section}

\newcommand{\be}{\begin{equation}}
\newcommand{\ee}{\end{equation}}

\newcommand{\txpf}{\texorpdfstring}

\newcommand*\cc[1]{
\hspace{2pt}\begin{tikzpicture}[baseline=-3.5pt]
\node[draw,circle,inner sep=1pt] at (0,0) {\scriptsize $#1$};
\node at (0,-0.25) {\scriptsize};
\end{tikzpicture}\hspace{2pt} }
\newcommand*\cct[2]{
\hspace{-1pt}\begin{tikzpicture}[baseline=-3.5pt]
\node[draw,circle,inner sep=1pt] at (0,0) {\scriptsize $#1$};
\node at (0,-0.25) {\scriptsize $#2$};
\end{tikzpicture}\hspace{-1pt} }
\newcommand{\ointc}{\oint}
\newcommand{\pa}{\partial}
\newcommand{\la}{\leftarrow}
\newcommand{\ra}{\rightarrow}
\newcommand{\ua}{\uparrow}

\newcommand{\spm}{{\pm}}

\newcommand{\as}{\operatorname{arcsinh}}
\newcommand{\qh}{\sqrt{1-q^2}}
\newcommand{\qp}{\sqrt{1+q}}
\newcommand{\qm}{\sqrt{1-q}}
\newcommand{\BES}{\text{BES}}
\newcommand{\AFS}{\text{AFS}}
\newcommand{\HL}{\text{HL}}

\newcommand{\Jc}{\mathcal{J}}
\newcommand{\cl}{\ell}
\newcommand{\IndA}{M}
\newcommand{\IndB}{N}
\newcommand{\IndC}{P}
\newcommand{\IndD}{Q}
\newcommand{\IndE}{R}
\newcommand{\IndF}{S}

\newcommand{\hh}{\text{h}}
\newcommand{\rmp}{\text{p}}
\newcommand{\rmq}{\text{q}}
\newcommand{\rml}{\text{l}}
\newcommand{\rmP}{\text{P}}
\newcommand{\rmk}{\text{k}}
\renewcommand{\a}{\alpha}

\renewcommand{\b}{\beta}

\renewcommand{\c}{\gamma}

\newcommand{\g}{\gamma}

\newcommand{\e}{\epsilon}

\newcommand{\q}{\theta}

\newcommand{\n}{\nu}

\newcommand{\p}{\pi}
\newcommand{\vp}{\varpi}

\newcommand{\s}{\sigma}

\newcommand{\vf}{\varphi}

\begin{document}

\thispagestyle{empty}

\vspace{ -3cm} \thispagestyle{empty} \vspace{-1cm}
\begin{flushright}
\footnotesize
HU-EP-14/18\\
\end{flushright}

\begingroup\centering
{\Large\bfseries
$\mathbf{AdS_3 \times S^3 \times M^4}$ string S-matrices from unitarity cuts
\par}
\vspace{7mm}

\begingroup
Lorenzo~Bianchi and Ben~Hoare\\
\endgroup
\vspace{8mm}
\begingroup\small
\emph{Institut f\"ur Physik,
Humboldt-Universit\"at
zu Berlin\\ Newtonstra{\ss}e 15, 12489 Berlin, Germany}\\
\endgroup

\vspace{0.3cm}
\begingroup\small
{\tt $\{$lorenzo.bianchi, ben.hoare$\}$@\,physik.hu-berlin.de}
\endgroup
\vspace{1.0cm}

\textbf{Abstract}\vspace{5mm}\par
\begin{minipage}{14.7cm}

Continuing the program initiated in arXiv:1304.1798 we investigate unitarity
methods applied to two-dimensional integrable field theories. The one-loop
computation is generalized to encompass theories with different masses in the
asymptotic spectrum and external leg corrections. Additionally, the
prescription for working with potentially singular cuts is modified to cope
with an ambiguity that was not encountered before. The resulting methods are
then applied to three light-cone gauge string theories; i) $AdS_3 \times S^3
\times T^4$ supported by RR flux, ii) $AdS_3 \times S^3 \times S^3 \times S^1$
supported by RR flux and iii) $AdS_3 \times S^3 \times T^4$ supported by a mix
of RR and NSNS fluxes. In the first case we find agreement with the exact
result following from symmetry considerations and in the second case with
one-loop semiclassical computations. This agreement crucially includes the
rational terms and hence supports the conjecture that S-matrices of integrable
field theories are cut-constructible, up to a possible shift in the coupling.
In the final case, under the assumption that our methods continue to give all
rational terms, we give a conjecture for the one-loop phases.

\end{minipage}\par
\endgroup

\newpage

\tableofcontents

\newpage

\section{Introduction}

In \cite{Bianchi:2013nra,Engelund:2013fja} unitarity methods
\cite{Bern:1994zx,Bern:1994cg,Roib_Review} were applied to various
two-dimensional field theories with the aim of computing the two-particle to
two-particle S-matrix. One of the intriguing consequences of these works was
the observation that this approach is particularly powerful when applied to
integrable field theories.\footnote{S-matrices of integrable field theories are
in general rather special as they are heavily constrained by a set of physical
requirements -- no particle production, equality of the sets of incoming and
outgoing momenta and the factorization of the $n$-particle scattering amplitude
into a product of two-particle S-matrices
\cite{Zamolodchikov:1978xm,Dorey:1996gd}.\vspace{2pt}} In \cite{Bianchi:2013nra} it was
observed that the one-loop S-matrices, including rational terms, for a number
of integrable theories (including sine-Gordon and various generalizations) are
completely cut-constructible (up to possible finite shifts in the coupling). In
particular, the unitarity construction automatically accounts for additional
contributions from one-loop counterterms that are required to preserve
integrability in the standard Feynman diagram computation. Furthermore, as the
unitarity construction reduces the one-loop computation to scalar bubble
integrals, which are finite in two dimensions, issues of regularization are
bypassed.

The motivation for investigating these methods in the context of
two-dimensional integrable field theories is their potential use in the study
of string backgrounds with worldsheet integrability. The classic example is the
$AdS_5 \times S^5$ superstring \cite{Metsaev:1998it}, whose worldsheet theory
was demonstrated to be classically integrable in \cite{Bena:2003wd}. In
\cite{Bianchi:2013nra} it was shown that the one-loop unitarity computation of
the light-cone gauge S-matrix, the tree-level components of which were computed
in \cite{Klose:2006zd}, matches the expansion of the exact S-matrix
\cite{Beisert:2005tm,Beisert:2006ez}, including rational terms. The exact
S-matrix was derived as a result of an extensive body of work, following from
considerations of symmetries, integrability and perturbation theory in both
gauge and string theory. For further details and references see the reviews
\cite{Arutyunov:2009ga,Beisert:2010jr}. In \cite{Engelund:2013fja} the one- and
two-loop logarithmic terms were also shown to be in agreement. In general, in an
integrable theory the logarithmic terms are required to exponentiate into a set
of overall phase factors. This was confirmed to be the case for a variety of
integrable string backgrounds of relevance in the context of the AdS/CFT
correspondence \cite{Engelund:2013fja,Abbott:2013kka}.

In this paper we will develop and apply the methods of \cite{Bianchi:2013nra}
to a class of integrable theories that arise as the light-cone gauge-fixing
\cite{Frolov:2006cc} of $AdS_3 \times S^3 \times M^4$ string backgrounds
\cite{Pesando:1998wm,Rahmfeld:1998zn,Babichenko:2009dk,Zarembo:2010sg,Cagnazzo:2012se}.
We will focus on the following three cases. The first is the simplest and is
when the compact manifold is $T^4$ with the background supported by RR flux.
The second is when the compact manifold is $S^3 \times S^1$, again supported by
RR flux. For the last we return to $T^4$, but with the background now supported
by a mix of RR and NSNS fluxes. The final two can both be thought of (at least
at the level of the decompactified light-cone gauge worldsheet theory) as
deformations of the first. For the $S^3 \times S^1$ theory we have a parameter
$\alpha \in [0,1]$, such that for $\a = 0$ and $1$, one or other of the two
three-spheres blows up, while in the mixed flux theory we have a parameter $q
\in [0,1]$, where $q=0$ corresponds to pure RR flux and $q=1$ to pure NSNS
flux.

The S-matrix of interest to us describes the scattering of excitations on the
decompactified string worldsheet in the uniform light-cone gauge
\cite{Frolov:2006cc}. The masses of the asymptotic excitations are given by the
expansion around the BMN string \cite{Berenstein:2002jq}. For the theories
under consideration we have the following spectra
\renewcommand{\arraystretch}{1.5}
\begin{center}
\begin{tabular}{ll}
\hline\hline Theory & Spectrum
\\ \hline\hline
$AdS_3 \times S^3 \times T^4$ (RR flux) & $(4+4) \times 1 \qquad (4+4) \times 0$
\\ \hline
$AdS_3 \times S^3 \times S^3 \times S^1$ (RR flux) \qquad\qquad &
$(2+2) \times 1 \qquad (2+2) \times \alpha $ \\ & $\quad (2+2) \times 1-\alpha \qquad (2+2) \times 0$
\\ \hline
$AdS_3 \times S^3 \times T^4$ (mixed flux) & $(4+4) \times \sqrt{1-q^2} \qquad (4+4) \times 0$
\\ \hline\hline \end{tabular}
\end{center}
where $(n+n)$ denotes bosons+fermions. As expected, in each case we have
$(8+8)$ excitations in total and the masses of the bosons match those of the
fermions. All three cases feature massless modes, which need careful treatment
in two dimensions. In the main text we will argue that if we restrict to
massive external legs, then we can ignore the massless modes completely in the
one-loop unitarity computation. Therefore in this paper we will do so, leaving
the question of massless modes in two-dimensional unitarity computations for
future work.

The main input of the one-loop unitarity computation is the tree-level S-matrix
of the theory. Various components of the tree-level S-matrices for the $T^4$
and $S^3 \times S^1$ backgrounds supported by RR flux were computed in
\cite{Sundin:2013ypa,Rughoonauth:2012qd}, and in \cite{Hoare:2013pma} for the
mixed flux case. These results, along with the symmetries and integrability of
the theory, can be used to completely determine the tree-level S-matrix. For
the $AdS_3 \times S^3 \times S^3 \times S^1$ background we will also require
additional input. In this case the expansion of the light-cone gauge-fixed
Lagrangian contains odd powers of the fields, in particular cubic terms
\cite{Sundin:2014sfa,Sundin:2012gc}. This, along with the asymptotic mass
spectrum, implies that there can be external leg corrections contributing to
the one-loop unitarity computation. We deal with these additional terms by
considering a unitarity computation involving form factors (i.e. partially
off-shell quantities) on either side of the cut.

In \cite{Engelund:2013fja} the one- and two-loop logarithmic terms were studied
for these theories. Therefore, our main aim here is to compute the rational
terms. The motivation for this is two-fold. First, we would like to demonstrate
that the matrix structure (that is the S-matrix up to overall phase factors)
matches that which is found via symmetries and integrability
\cite{Borsato:2012ud,Borsato:2012ss,Borsato:2013qpa,Borsato:2013hoa,Hoare:2013ida,Hoare:2013lja}.
Second, we would like to investigate the phases of these theories, including
the rational terms.  Thus far there is only an all-loop conjecture (supported
by semiclassical one-loop computations in
\cite{Abbott:2012dd,Beccaria:2012kb}\footnote{The two semiclassical one-loop 
computations \cite{Abbott:2012dd,Beccaria:2012kb} are not in complete
agreement. While the logarithmic terms match, the rational terms we find from
unitarity methods agree with those in \cite{Abbott:2012dd} and the expansion of
the exact result \cite{Borsato:2013hoa}. It is currently unclear what the
precise reason for the disagreement between \cite{Abbott:2012dd} and
\cite{Beccaria:2012kb} is.\label{clarification}}) for the phases in the $AdS_3
\times S^3 \times T^4$ case supported by RR flux \cite{Borsato:2013hoa}.  There
is also a semiclassical one-loop computation of the phases in the $AdS_3 \times
S^3 \times S^3 \times S^1$ case in \cite{Abbott:2013ixa}. In these two models we
check that our methods reproduces the known results at one-loop, providing further
support for the conjecture of \cite{Bianchi:2013nra} that the S-matrices of
integrable theories are cut-constructible (up to possible shifts in the
coupling).  Given this, we can therefore use unitarity techniques to conjecture
one-loop expressions for the phases in the mixed flux case, which should then
provide an insight into how to deform the all-loop phases of
\cite{Borsato:2013hoa}.

We shall start in section \ref{sec:gp} with an outline of the general method,
emphasizing the modifications to the construction in \cite{Bianchi:2013nra}
that are required to deal with the three theories of interest. There are three
such modifications -- a prescription for the rational t-channel contribution
when the consistency condition of \cite{Bianchi:2013nra} is not satisfied, a
discussion of what happens with particles of different mass in the asymptotic
spectrum and the possibility of external leg corrections. In the following
sections the procedure is then applied to the $AdS_3 \times S^3 \times M^4$
string backgrounds.

\section{General principles}\label{sec:gp}

In \cite{Bianchi:2013nra} unitary methods were used to the construct the
one-loop S-matrix of various two-dimensional integrable field theories.
Agreement with known results was found not only for the logarithmic terms, but
also with the rational part (up to shifts in the coupling). The construction of
the rational part was based on a prescription that dealt with various singular
cuts that arise when applying the usual unitarity methods. This prescription
required that a consistency condition on the tree-level S-matrix was satisfied.
However, for various other examples of interest in the context of integrable
string backgrounds, this consistency condition is not satisfied. In this
section we will outline the construction of \cite{Bianchi:2013nra} (see also
\cite{Engelund:2013fja} for a discussion of the logarithmic terms to two loops)
and describe a modification of the prescription that will allow the method to
be used when the consistency condition is not satisfied.

\subsection{Theories with a single mass}\label{singlemass}

The object of interest is the two-particle S-matrix, defined in terms of the
four-point amplitude
\begin{equation}\label{eqn:4ptamp}
\langle\Phi^\IndC(\rmq)\Phi^\IndD(\rmq')\,|\mathbb{S}|\,\Phi_\IndA(\rmp)\Phi_\IndB(\rmp')\rangle=\mathcal{A}_{\IndA\IndB}^{\IndC\IndD}(\rmp,\rmp',\rmq,\rmq')~.
\end{equation}
Here $\mathbb{S}$ is the scattering operator, $\IndA,\IndB,\ldots$ are indices
running over the particle content of the theory and $\rmp,\rmp',\rmq,\rmq'$ are
the on-shell momenta of the fields. For now we will restrict to the case where
all the particles have equal non-vanishing mass, which we set to one. As a
consequence of momentum conservation, the four-point amplitude takes the form
\begin{equation}\label{eqn:ampcons}
\mathcal{A}_{\IndA\IndB}^{\IndC\IndD}(\rmp,\rmp',\rmq,\rmq')=(2\pi)^2 \delta^{(2)}(\rmp+\rmp'-\rmq-\rmq')\, \widetilde{\mathcal{A}}_{\IndA\IndB}^{\IndC\IndD}(\rmp,\rmp',\rmq,\rmq')~.
\end{equation}
Furthermore, at four points, two-dimensional kinematics implies that the set of
initial momenta is preserved in the scattering process, as demonstrated by the
following identity
\begin{equation}\label{delta2d}
\delta^{(2)}(\rmp+\rmp'-\rmq-\rmq')=\frac{\Jc(p,p')}{4\e\e'}\,\big(2\e\,\delta(p - q)\,2\e'\delta(p' - q') + 2\e\, \delta(p - q')\,2\e'\delta(p' - q) \big)~,
\end{equation}
where $p,p',q,q'$ are the spatial momentum and the Jacobian
$\Jc(p,p')=1/(\partial \e/\partial p-\partial\e'/\partial p')$ depends on the
on-shell energies $\e(p),\e'(p')$. Note that we have assumed the particle
velocities are ordered as $v = \partial \e/\partial p > \partial \e'/\partial
p'= v'$ and for the spatial momentum $\delta$-functions we have used a
normalization that becomes the standard Lorentz-invariant one in the
relativistic case.

Substituting \eqref{delta2d} into \eqref{eqn:ampcons} we find two terms.
Without loss of generality we can consider just the amplitude associated to the
first product of $\delta$-functions, $2\e\, \delta(p - q)\,2\e'\delta(p' - q')$.
The two-particle S-matrix is then defined as
\begin{equation}\label{AandS}
S_{\IndA\IndB}^{\IndC\IndD}(p,p')\equiv \frac{\Jc(p,p')}{4\e\e'}\widetilde{\mathcal{A}}_{\IndA\IndB}^{\IndC\IndD}(\rmp,\rmp',\rmp,\rmp')~,
\end{equation}
As usual we can expand in powers of the coupling
\begin{equation}
S_{\IndA\IndB}^{\IndC\IndD}(p,p') = \delta_{\IndA}^{\IndC}\delta_{\IndB}^{\IndD} + i h^{-1} T^{(0)}{}_{\IndA\IndB}^{\IndC\IndD}(p,p') +
i h^{-2} T^{(1)}{}_{\IndA \IndB}^{\IndC \IndD}(p,p') + \mathcal{O}(h^{-3})\ ,
\end{equation}
with the identity at leading order, followed by the tree-level S-matrix
$T^{(0)}$ at order $h^{-1}$ and the $n$-loop S-matrix $T^{(n)}$ at order
$h^{-n-1}$.

In this paper we will be interested in computing the cut-constructible part of
$T^{(1)}$ from the tree-level S-matrix $T^{(0)}$ for the light-cone gauge
S-matrices for string theories in $AdS_3 \times S^3 \times M^4$ backgrounds.
For these theories the inverse string tension plays the role of the coupling
$h^{-1}$. It is well known that the dispersion relation is in general
non-relativistic, however, at quadratic order in the near-BMN expansion (i.e.
for the free states in perturbation theory) it is. Therefore, for our
purposes the pre-factor in \eqref{AandS}, i.e. the contribution coming from the
Jacobian, is just given by
\begin{equation}\label{jdef}
J(p,p') = \frac{1}{4(e'p-e p')} \ , \qquad e = \sqrt{p^2 + 1} \ , \qquad e' = \sqrt{p'^2 + 1} \ .
\end{equation}
In general we will use $e$ to denote the energies of the free states, and
$\e$ to denote their all-order form.

\begin{figure}[t]
\begin{center}
\begin{tikzpicture}[line width=1pt,scale=1.3,baseline = -250]
\draw[-] (-5,-3) -- (-4,-4);
\draw[-] (-5,-5) -- (-4,-4);
\draw[-] (-2,-4) -- (-1,-3);
\draw[-] (-2,-4) -- (-1,-5);
\draw (-3,-4) circle (1cm);
\draw[|-|,dashed,blue!70,line width=1pt] (-3,-2.5) -- (-3,-5.5);
\draw[->] (-4.7,-3.0) -- (-4.3,-3.4);
\node at (-4.7,-2.8) {$\rmp$};
\draw[->] (-4.7,-5.0) -- (-4.3,-4.6);
\node at (-4.7,-5.2) {$\rmp'$};
\draw[<-] (-1.3,-3.0) -- (-1.7,-3.4);
\node at (-1.3,-2.8) {$\rmq'$};
\draw[<-] (-1.3,-5.0) -- (-1.7,-4.6);
\node at (-1.3,-5.2) {$\rmq$};
\draw[<-] (-3.1,-3.2) -- (-3.5,-3.35);
\node at (-3.2,-3.5) {$\rml_1$};
\draw[->] (-2.9,-4.8) -- (-2.5,-4.65);
\node at (-2.8,-4.5) {$\rml_2$};
\node at (-3.4,-2.85) {$\IndE$};
\node at (-2.6,-5.15) {$\IndF$};
\node at (-5.2,-3) {$\IndA$};
\node at (-5.2,-5) {$\IndB$};
\node at (-0.8,-5) {$\IndC$};
\node at (-0.8,-3) {$\IndD$};
\node [circle,draw=black!100,fill=black!5,thick,opacity=.95] at (-4,-4) {\footnotesize$\mathcal{A}^{(0)}$\normalsize};
\node [circle,draw=black!100,fill=black!5,thick,opacity=.95] at (-2,-4) {\footnotesize$\mathcal{A}^{(0)}$\normalsize};
\end{tikzpicture}
\begin{tikzpicture}[line width=1pt,scale=1.5,rotate=90]
\draw[-] (-5,-3) -- (-4,-4);
\draw[-] (-5,-5) -- (-4,-4);
\draw[-] (-2,-4) -- (-1,-3);
\draw[-] (-2,-4) -- (-1,-5);
\draw (-3,-4) circle (1cm);
\draw[|-|,dashed,blue!70,line width=1pt] (-3,-2.5) -- (-3,-5.5);
\draw[->] (-4.7,-3.0) -- (-4.3,-3.4);
\node at (-4.7,-2.8) {$\rmp'$};
\draw[<-] (-4.7,-5.0) -- (-4.3,-4.6);
\node at (-4.7,-5.2) {$\rmq'$};
\draw[->] (-1.3,-3.0) -- (-1.7,-3.4);
\node at (-1.3,-2.8) {$\rmp$};
\draw[<-] (-1.3,-5.0) -- (-1.7,-4.6);
\node at (-1.3,-5.2) {$\rmq$};
\draw[<-] (-3.1,-3.2) -- (-3.5,-3.35);
\node at (-3.2,-3.5) {$\rml_1$};
\draw[<-] (-2.9,-4.8) -- (-2.5,-4.65);
\node at (-2.8,-4.5) {$\rml_2$};
\node at (-3.4,-2.85) {$\IndE$};
\node at (-2.6,-5.15) {$\IndF$};
\node at (-5.2,-3) {$\IndB$};
\node at (-5.2,-5) {$\IndD$};
\node at (-0.8,-5) {$\IndC$};
\node at (-0.8,-3) {$\IndA$};
\node [circle,draw=black!100,fill=black!5,thick,opacity=.95] at (-4,-4) {\footnotesize$\mathcal{A}^{(0)}$\normalsize};
\node [circle,draw=black!100,fill=black!5,thick,opacity=.95] at (-2,-4) {\footnotesize$\mathcal{A}^{(0)}$\normalsize};
\end{tikzpicture}
\hspace{10pt}
\begin{tikzpicture}[line width=1pt,scale=1.5,rotate=90]
\draw[-] (-5,-3) -- (-4,-4);
\draw[-] (-5,-5) -- (-4,-4);
\draw[-] (-2,-4) -- (-1,-3);
\draw[-] (-2,-4) -- (-1,-5);
\draw (-3,-4) circle (1cm);
\draw[|-|,dashed,blue!70,line width=1pt] (-3,-2.5) -- (-3,-5.5);
\draw[->] (-4.7,-3.0) -- (-4.3,-3.4);
\node at (-4.7,-2.8) {$\rmp'$};
\draw[<-] (-4.7,-5.0) -- (-4.3,-4.6);
\node at (-4.7,-5.2) {$\rmq$};
\draw[->] (-1.3,-3.0) -- (-1.7,-3.4);
\node at (-1.3,-2.8) {$\rmp$};
\draw[<-] (-1.3,-5.0) -- (-1.7,-4.6);
\node at (-1.3,-5.2) {$\rmq'$};
\draw[<-] (-3.1,-3.2) -- (-3.5,-3.35);
\node at (-3.2,-3.5) {$\rml_1$};
\draw[<-] (-2.9,-4.8) -- (-2.5,-4.65);
\node at (-2.8,-4.5) {$\rml_2$};
\node at (-3.4,-2.85) {$\IndE$};
\node at (-2.6,-5.15) {$\IndF$};
\node at (-5.2,-3) {$\IndB$};
\node at (-5.2,-5) {$\IndC$};
\node at (-0.8,-5) {$\IndD$};
\node at (-0.8,-3) {$\IndA$};
\node [circle,draw=black!100,fill=black!5,thick,opacity=.95] at (-4,-4) {\footnotesize$\mathcal{A}^{(0)}$\normalsize};
\node [circle,draw=black!100,fill=black!5,thick,opacity=.95] at (-2,-4) {\footnotesize$\mathcal{A}^{(0)}$\normalsize};
\end{tikzpicture}
\caption{Diagrams representing s-, t- and u-channel cuts contributing to the four-point one-loop amplitude.}
\label{stu}\nonumber
\end{center}
\end{figure}
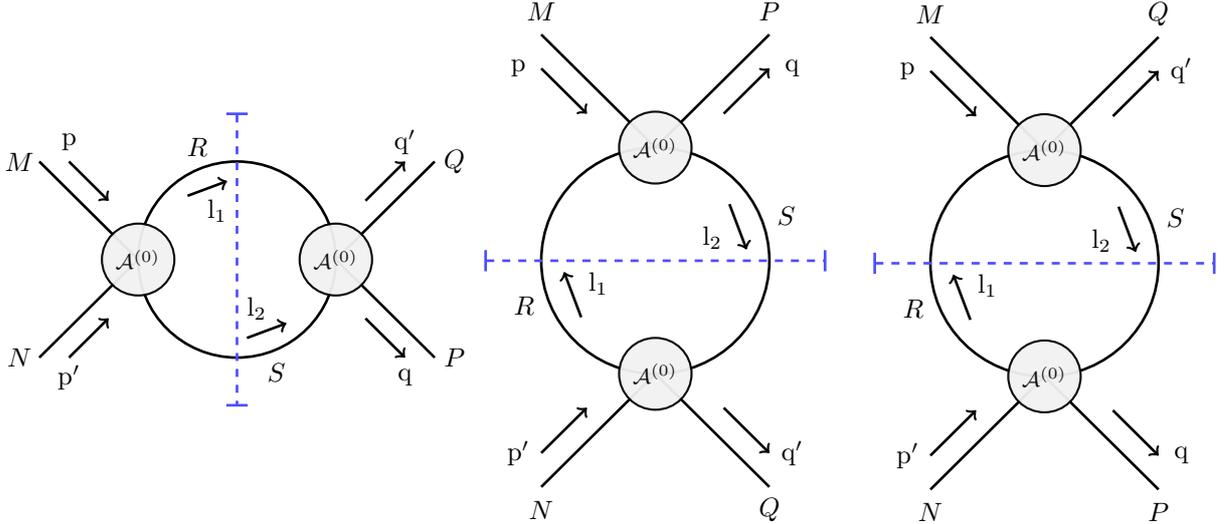

In general, there are three possible contributions that can arise in a
unitarity computation, which are shown in figure~\ref{stu}. We ignore tadpoles
and graphs built from a three- and five-point amplitude. In the standard
unitarity procedure such graphs have no physical two-particle cuts and
therefore they can safely be ignored. However in higher dimensions, a recipe to
deal with tadpole diagrams in the context of generalized unitarity for massive
theories was given in \cite{Britto:2010um}. In two dimensions the situation is
slightly different.
In particular, tadpole diagrams require the introduction of a regularization
since they develop a logarithmic divergence.
Since our procedure is inherently finite it is not clear
how tadpoles should be included, but it appears that they do not need to be to
construct the one-loop S-matrix (up to possible shifts in the coupling) as we
have explicitly checked in all the cases under consideration. We leave further
analysis of tadpole graphs for future studies.

Here we will outline the derivation (for details the reader should refer to
\cite{Bianchi:2013nra}) of the one-loop S-matrix following from unitarity
methods. The key observation is that as we have two-particle cuts in a
two-dimensional field theory this completely freezes the momentum. Therefore,
using this the tree-level amplitudes on either side of the cut can be pulled
out of the loop integral, with the loop momenta taking their frozen values.
After doing this we return the internal on-shell propagators off-shell leaving
us with scalar bubble integrals
\begin{equation}\label{sbi}
I(\rmP^2,m,m')=\int \frac{d^2 \rmk}{(2\pi)^2} \frac{1}{(\rmk^2-m^2+i\e) ((\rmk-\rmP)^2-m'^2+i\e)}~,
\end{equation}
with coefficients given by contractions of tree-level amplitudes. With this in
mind it is useful to define the following tensor contractions
\begin{align}
(A \cc{s} B)_{\IndA\IndB}^{\IndC\IndD}(p,p') &= A_{\IndA\IndB}^{\IndE\IndF}(p,p')B_{\IndE\IndF}^{\IndC\IndD}(p,p')\ ,\label{scont}
\\ (A \cc{u} B)_{\IndA\IndB}^{\IndC\IndD}(p,p') &= (-1)^{([\IndC]+[\IndF])([\IndD]+[\IndE])}A_{\IndA\IndE}^{\IndF\IndD}(p,p')B_{\IndF\IndB}^{\IndC\IndE}(p,p')\ ,\label{ucont}
\\ (A \cct{t}{\la} B)_{\IndA\IndB}^{\IndC\IndD}(p,p') &= (-1)^{[\IndC][\IndF] + [\IndE][\IndF]} A_{\IndA\IndE}^{\IndF\IndC}(p,p) B_{\IndF\IndB}^{\IndE\IndD}(p,p')\ ,\label{tcont1}
\\ (A \cct{t}{\ra} B)_{\IndA\IndB}^{\IndC\IndD}(p,p') &= (-1)^{[\IndD][\IndE] + [\IndE][\IndF]} A_{\IndA\IndE}^{\IndC\IndF}(p,p') B_{\IndF\IndB}^{\IndD\IndE}(p',p')\ ,\label{tcont2}
\end{align}
where $[\IndA] = 0$ for a boson and $1$ for a fermion, and the following scalar
bubble integrals
\begin{align}\label{ints}
I_s & \equiv I((\rmp + \rmp')^2,1,1) = \frac{1}{4(e'p-e p')}(1-\frac{\as(e'p-e p')}{i\pi}) = \frac{J}{i\pi}(i\pi-\theta) \ ,
\\\label{intt}
I_t & \equiv I(0,1,1) = \frac{1}{4\pi i} \ ,
\\\label{intu}
I_u & \equiv I((\rmp - \rmp')^2,1,1) = \frac{1}{4(e'p-e p')}\frac{\as(e'p-e p')}{i\pi} = \frac{J\theta}{i\pi} \ ,
\end{align}
where we have used \eqref{jdef} and defined
\begin{equation}\label{thetadef}
\theta \equiv \as(e'p-e p')\ .
\end{equation}

The final step is to set $q = p$ and $q' = p'$ to extract the one-loop S-matrix
\begin{equation}
T^{(1)}= \frac {iJ}2 (C_s I_s + C_t I_t + C_u I_u) \ ,
\end{equation}
where for clarity we have suppressed the indices. Here $J$ is the contribution
from the Jacobian and the $\tfrac12$ is the symmetry factor. The matrices
$C_{s,u}$ are given by
\begin{align}
C_{s} &= \widetilde T^{(0)} \cc{s} \widetilde T^{(0)} \ ,
\qquad
C_{u} = \widetilde T^{(0)} \cc{u} \widetilde T^{(0)} \ ,
\end{align}
where
\begin{equation}\label{ttttt}
\widetilde T^{(0)}=J^{-1}T^{(0)} \ .
\end{equation}

The t-channel contraction is more subtle as there two possible choices for
freezing the loop momenta (i.e. in terms of $p$ and $q$ or $p'$ and $q'$) giving
potentially different results. These correspond to the two contractions in
eqs.~\eqref{tcont1} and \eqref{tcont2}. In the theories considered in
\cite{Bianchi:2013nra} it turned out that these two choices always gave the
same result, i.e.
\begin{equation}
\widetilde T^{(0)} \cct{t}{\la} \widetilde T^{(0)}=\widetilde T^{(0)} \cct{t}{\ra} \widetilde T^{(0)}. \label{cons}
\end{equation}
and this issue could safely be ignored. However, for the S-matrices we will
consider in this paper, i.e. the light-cone gauge S-matrices for strings in
$AdS_3 \times S^3 \times M^4$ this is no longer the case. In particular we note
that the function $\widetilde T^{(0)}(p,p)$ cannot have any momentum dependence
in a relativistic theory,\footnote{Let us recall that in a relativistic theory
the S-matrix depends only on the difference of rapidities, which vanishes for
$p'=p$.\vspace{2pt}} whereas in a non-relativistic theory it can depend on $p$,
generating an asymmetry between $p$ and $p'$.\footnote{In
\cite{Bianchi:2013nra} the only non-relativistic theory that was considered was
the light-cone gauge-fixed string in $AdS_5\times S^5$ for which, despite the
asymmetry, eq.~\eqref{cons} still holds.\vspace{2pt}} Hence it is natural to
conjecture that we should take the average of the two contractions.
Therefore
\begin{equation}
C_t=\frac12(\widetilde T^{(0)} \cct{t}{\la} \widetilde T^{(0)}+\widetilde T^{(0)} \cct{t}{\ra} \widetilde T^{(0)})\ .
\end{equation}
To conclude the construction we can use the explicit expressions of the
integrals $I_{s,t,u}$ in eqs.~\eqref{ints} to \eqref{intu} and the relation
between $T^{(0)}$ and $\widetilde T^{(0)}$ \eqref{ttttt} to rewrite the
one-loop result as
\begin{equation}\label{result}
T^{(1)}= \frac{\q}{2\p} (T^{(0)}\cc{u} T^{(0)}-T^{(0)} \cc{s} T^{(0)})+\frac i2 T^{(0)} \cc{s} T^{(0)} +\frac1{16\p} (\widetilde T^{(0)} \cct{t}{\la} T^{(0)}+T^{(0)} \cct{t}{\ra} \widetilde T^{(0)})\ ,
\end{equation}
where, under the assumption that $T^{(0)}$ is real, there is a natural split of
the result into three pieces; a logarithmic part, an imaginary rational part,
and a real rational part.

\subsection{Theories with multiple masses}\label{2mass}

We will now generalize the above construction to the case where the asymptotic
spectrum contains particles of different mass. In this derivation we will
restrict to theories whose tree-level S-matrix is integrable, in particular,
using the consequence that the set of outgoing momenta is a permutation of the
set of incoming momenta. This means that, for the reasons explained in section
\ref{singlemass}, tadpoles and one-loop graphs built from a three- and
five-point amplitude will be ignored in the unitarity computation. Therefore we
are again left with the three contributions given in figure~\ref{stu}.

We consider the configuration in which the external legs with indices $M$ and
$P$ have mass $m$ and the associated momenta are equal ($p=q$) and $N$ and $Q$
have mass $m'$ with $p'=q'$.\footnote{Our procedure implies that if we assume
the set of outgoing momenta is equal to a permutation of the set of incoming
momenta at tree level, this property automatically extends to one loop.\vspace{2pt}}
For the s- and u-channels the story is then largely the same as the single-mass
case. It follows from the assumptions outlined in the previous paragraph that
when the two propagators are cut the internal loop momenta are frozen to the
values of the external momenta. The tree-level amplitudes on either side of the
cut can then be pulled out of the integral and we are left with scalar bubble
integrals with coefficients given by contractions of tree-level amplitudes.
Working through the remaining steps, which are essentially identical to the
single-mass case, it is clear that the contribution from these graphs is given
by
\begin{equation}\label{resultsumm}
T^{(1)}_{s,u} = \frac{\q}{2\p} (T^{(0)}\cc{u} T^{(0)}-T^{(0)} \cc{s} T^{(0)})+\frac i2 T^{(0)} \cc{s} T^{(0)}\ ,
\end{equation}
where
\begin{equation}\begin{split}
& \theta \equiv \as \big(\frac{e'p-e p'}{mm'}\big)\ , \qquad e = \sqrt{p^2 + m^2} \ , \qquad e' = \sqrt{p'^2 + m'^2} \ .
\\ & I_s \equiv I((\rmp + \rmp')^2,m,m') = \frac{1}{4(e'p-e p')}(1-\frac{\as(\frac{e'p-e p'}{mm'})}{i\pi}) = \frac{J}{i\pi}(i\pi-\theta) \ ,
\\ & I_u \equiv I((\rmp - \rmp')^2,m,m') = \frac{1}{4(e'p-e p')}\frac{\as(\frac{e'p-e p'}{mm'})}{i\pi} = \frac{J\theta}{i\pi} \ ,
\end{split}\end{equation}
Here $m$ and $m'$ are the masses of the two particles being scattered and the
scalar bubble integral $I(\rmP^2,m,m')$ is defined in eq.~\eqref{sbi}.
Eq.~\eqref{resultsumm} therefore fixes the logarithmic and imaginary rational
parts of the one-loop result.

The real rational part, which comes from the t-channel contribution, is, as
before, more subtle. In the single-mass case, the guiding principle for
computing the t-channel cuts was to only fix $q=p$ and $q'=p'$ at the end in
order to avoid ill-defined expressions in the intermediate steps. Therefore,
let us consider the t-channel graph in figure~\ref{stu} with the external legs
with indices $M$ and $P$ having mass $m$, $N$ and $Q$ mass $m'$ and the loop
legs mass $m_l$, but $p$, $q$, $p'$ and $q'$ kept arbitrary, i.e. we do
{\em not} fix $q = p$ and $q'=p'$.

After putting the loop legs on-shell the loop momenta are fixed by the momentum
conservation delta functions in terms of the external momenta. Solving
in terms of $\rmp$ and $\rmq$ we find
\begin{equation}\begin{split}\label{sol1}
l_{1\pm}^{\uparrow} = & \frac12\big[ q_\pm - p_\pm + \sqrt{(q_\pm-p_\pm)^2 + 4 \frac{m_l^2}{m^2} q_\pm p_\pm}\big] \ ,
\\
l_{2\pm}^{\uparrow} = & \frac12\big[ p_\pm - q_\pm + \sqrt{(p_\pm-q_\pm)^2 + 4 \frac{m_l^2}{m^2} p_\pm q_\pm}\big]\ ,
\end{split}\end{equation}
while solving in terms of $\rmp'$ and $\rmq'$ gives
\begin{equation}\begin{split}\label{sol2}
l_{1\pm}^{\downarrow} = & \frac12\big[ p'_\pm - q'_\pm + \sqrt{(p'_\pm-q'_\pm)^2 + 4 \frac{m_l^2}{m^2} p'_\pm q'_\pm}\big]\ ,
\\
l_{2\pm}^{\downarrow} = & \frac12\big[ q'_\pm - p'_\pm + \sqrt{(q'_\pm-p'_\pm)^2 + 4 \frac{m_l^2}{m'^2} q'_\pm p'_\pm}\big] \ ,
\end{split}\end{equation}
where the light-cone momenta are defined in appendix \ref{notations}. The
first solution \eqref{sol1} then gives a contribution proportional to
\begin{equation}\label{expr1}
(-1)^{[P][S]+[R][S]}\widetilde{\mathcal{A}}^{(0)}{}_{MR}^{SP}(\rmp,\rml_1^\ua,\rml_2^\ua,\rmq)
\widetilde{\mathcal{A}}^{(0)}{}_{SN}^{RQ}(\rml_2^\ua,\rmp',\rml_1^\ua,\rmq')\ .
\end{equation}
The arguments of the second factor of $\widetilde{\mathcal{A}}^{(0)}$ contain
all four of the external momenta and therefore this part is well-defined when
we fix $q=p$ and $q'=p'$. Therefore, let us focus on the first factor of
$\widetilde{\mathcal{A}}^{(0)}$, whose arguments only depend on two of the
momenta. Recalling that in an integrable theory the amplitude should vanish
unless the set of outgoing momenta is a permutation of the set of incoming
momenta, it follows that this first factor vanishes unless $m_l = m$. In this
case \eqref{expr1} reduces to
\begin{equation}
(-1)^{[P][S]+[R][S]}\widetilde{\mathcal{A}}^{(0)}{}_{MR}^{SP}(\rmp,\rmq,\rmp,\rmq)
\widetilde{\mathcal{A}}^{(0)}{}_{SN}^{RQ}(\rmp,\rmp',\rmq,\rmq')\ .
\end{equation}
Finally setting $q=p$ and $q'=p'$ this expression can then be written in terms
of tree-level S-matrices. A similar logic follows for the second solution
\eqref{sol2}, except that here the contribution vanishes unless $m_l = m'$.

It therefore follows that the contribution from the t-channel is given by
\begin{equation}\label{resultsummt}
T^{(1)}_t = \frac1{16\p}( \frac1{m^2} \widetilde T^{(0)} \cct{t}{\la} T^{(0)}+ \frac{1}{m'^2}T^{(0)} \cct{t}{\ra} \widetilde T^{(0)})\ ,
\end{equation}
where $\widetilde T^{(0)}$ in the first term is built from the tree-level
S-matrix for the scattering of two excitations of mass $m$, while in the second
term it is built from the tree-level S-matrix for two excitations of mass $m'$.
We have included an additional factor of $\tfrac12$ as we should still use both
vertices to solve for the loop momenta and take the average.

Combining eqs.~\eqref{resultsumm} and \eqref{resultsummt} we find that the
one-loop result in the case where an excitation of mass $m$ is scattered with
an excitation of mass $m'$ is given by
\begin{equation}\label{resulttwo}
T^{(1)}= \frac{\q}{2\p} (T^{(0)}\cc{u} T^{(0)}-T^{(0)} \cc{s} T^{(0)})
+ \frac i2 T^{(0)} \cc{s} T^{(0)}
+ \frac1{16\p} (\frac{1}{m^2}\widetilde T^{(0)} \cct{t}{\la} T^{(0)}+\frac{1}{m'^2}T^{(0)} \cct{t}{\ra} \widetilde T^{(0)})\ ,
\end{equation}
where, again under the assumption that $T^{(0)}$ is real, there is a natural
split of the result into three pieces; a logarithmic part, an imaginary
rational part, and a real rational part. Setting $m=m'=1$ we see that this
formula reduces to, and hence incorporates, the single-mass case given in
eq.~\eqref{result}.

A key consequence of the results in this section is that the cut-constructible
one-loop S-matrix for the scattering of a particle of mass $m$ with one of mass
$m'$ is built from the corresponding tree-level S-matrix along with the
tree-level S-matrices for the scattering of two particles of mass $m$ and for
two particles of mass $m'$, both evaluated at equal momenta. In particular
there are no contributions containing tree-level S-matrices for particles of
masses other than $m$ and $m'$. This will be important in later sections as it
allows us to construct the one-loop cut-constructible S-matrix for various
sectors without knowing the full tree-level S-matrix.

The result \eqref{resulttwo} deserves a comment regarding its relation to
integrability and the Yang-Baxter equation. The Yang-Baxter equation is a cubic
matrix equation that should be satisfied by S-matrices describing scattering in
integrable theories. Up to signs related to fermions, which we are not
concerned with for this schematic discussion, it can be written as
\begin{equation}
\mathbb{S}_{12} \mathbb{S}_{13} \mathbb{S}_{23} =
\mathbb{S}_{23} \mathbb{S}_{13} \mathbb{S}_{12} \ ,
\end{equation}
where these operators are acting on a three-particle state and the indices
denote the particles that are being scattered. The first non-trivial order in
its perturbative expansion is called the classical Yang-Baxter equation and is
a relation that is quadratic in the tree-level S-matrix,
\begin{equation}\label{cybe}
[\mathbb{T}_{12}^{(0)},\mathbb{T}^{(0)}_{13}] +
[\mathbb{T}_{12}^{(0)},\mathbb{T}^{(0)}_{23}] +
[\mathbb{T}_{13}^{(0)},\mathbb{T}^{(0)}_{23}] = 0 \ .
\end{equation}
At the next order we find the following relation
\begin{equation}\begin{split}\label{olybe}
&[\mathbb{T}_{12}^{(0)},\mathbb{T}^{(1)}_{13}] +
[\mathbb{T}_{12}^{(0)},\mathbb{T}^{(1)}_{23}] +
[\mathbb{T}_{13}^{(0)},\mathbb{T}^{(1)}_{23}] -
[\mathbb{T}_{13}^{(0)},\mathbb{T}^{(1)}_{12}] -
[\mathbb{T}_{23}^{(0)},\mathbb{T}^{(1)}_{12}] -
[\mathbb{T}_{23}^{(0)},\mathbb{T}^{(1)}_{13}] =
\\
&\hspace{275pt}\mathbb{T}^{(0)}_{23} \mathbb{T}^{(0)}_{13} \mathbb{T}^{(0)}_{12} -
\mathbb{T}^{(0)}_{12} \mathbb{T}^{(0)}_{13} \mathbb{T}^{(0)}_{23} \ .
\end{split}\end{equation}
One can check that, assuming that the tree-level S-matrix satisfies the
classical Yang-Baxter equation \eqref{cybe}, the rational s-channel
contribution to the cut-constructible one-loop S-matrix precisely cancels the
terms cubic in the tree-level S-matrix on the right-hand side of
eq.~\eqref{olybe}. Therefore, for the one-loop cut-constructible S-matrix to
respect integrability the remaining terms should satisfy \eqref{olybe} with the
right-hand side set to zero. In general, this condition is not easy to solve,
but two solutions are clear. The first is the tree-level S-matrix itself (which
amounts to a shift in the coupling), and the second is any contribution
that can be absorbed into the overall phase factors.

It will turn out that of the three theories we are interested in, two satisfy
this property. For the $AdS_3 \times S^3 \times S^3 \times S^1$ background, the
one-loop cut-constructible S-matrix as defined by \eqref{resulttwo} has a
rational piece coming from the t-channel that does not satisfy \eqref{olybe}
with zero on the right-hand side. However, there is a meaning to these terms --
they are cancelled by corrections to the external legs, which we will now
discuss.

\subsection{External leg corrections}\label{sec:elc}

In the construction outlined thus far we have not included any discussion of
corrections to the external legs. As shall become apparent, for the $AdS_3
\times S^3 \times S^3 \times S^1$ background, these will be important even at
one loop. These corrections will give a rational contribution to the S-matrix
and can follow from the three types of Feynman diagrams in figure \ref{graphs}.
\begin{figure}[t]
\begin{center}
\begin{tikzpicture}[line width=1pt,scale=0.9]
\draw[-] (-5.4,-4) -- (-4,-4);
\draw[-] (-2,-4) -- (-0.6,-4);
\draw (-3,-4) circle (1cm);
\draw[->] (-5.2,-4.2) -- (-4.6,-4.2);
\node at (-5.2,-4.4) {$\rmp$};
\draw[->] (-1.4,-4.2) -- (-0.8,-4.2);
\node at (-1.4,-4.4) {$\rmp$};
\draw[<-] (-3.1,-3.2) -- (-3.5,-3.35);
\node at (-3.2,-3.6) {$\rml_1$};
\draw[<-] (-2.9,-4.8) -- (-2.5,-4.65);
\node at (-2.7,-4.4) {$\rml_2$};
\end{tikzpicture}
\qquad\qquad
\begin{tikzpicture}[line width=1pt,scale=0.9]
\draw[-] (-5.4,-5) -- (-0.6,-5);
\draw (-3,-4) circle (1cm);
\draw[->] (-5.2,-5.2) -- (-4.6,-5.2);
\node at (-5.2,-5.4) {$\rmp$};
\draw[->] (-1.4,-5.2) -- (-0.8,-5.2);
\node at (-1.4,-5.4) {$\rmp$};
\draw[<-] (-3.1,-3.2) -- (-3.5,-3.35);
\node at (-3.2,-3.6) {$\rml_{\hphantom{1}}$};
\end{tikzpicture}
\qquad\qquad
\begin{tikzpicture}[line width=1pt,scale=0.9]
\draw[-] (-5.4,-5) -- (-0.6,-5);
\draw[-] (-3,-5) -- (-3,-4.5);
\draw (-3,-3.75) circle (0.75cm);
\draw[->] (-5.2,-5.2) -- (-4.6,-5.2);
\node at (-5.2,-5.4) {$\rmp$};
\draw[->] (-1.4,-5.2) -- (-0.8,-5.2);
\node at (-1.4,-5.4) {$\rmp$};
\draw[<-] (-3.1,-3.2) -- (-3.45,-3.35);
\node at (-3.2,-3.6) {$\rml_{\hphantom{1}}$};
\end{tikzpicture}
\caption{Diagrams contributing to external leg corrections at one-loop.}
\label{graphs}
\end{center}
\end{figure}

We will be interested in external leg corrections at one loop that are caught
by unitarity. In order to approach this problem let us first review how
external leg corrections are usually dealt with in a standard Feynman diagram
calculation. We denote the sum of all one particle irreducible insertions into
a scalar propagator as $-i\Sigma(\rmp) = -i h^{-1}\Sigma^{(1)}(\rmp) +
\mathcal{O}(h^{-2})$, where $-i h^{-1}\Sigma^{(1)}(\rmp)$ is the one-loop
contribution. After re-summing one finds
\begin{equation}
\begin{tikzpicture}[line width=1pt,baseline=-3]
\draw[-] (-1,0)--(0,0);
\draw [fill=black!20] (0.5,0) circle (0.5);
\draw[-] (1,0)--(2,0);
\end{tikzpicture}
=\frac{i}{\rmp^2-m^2-\Sigma(\rmp)}
\end{equation}
Expanding $\Sigma(\rmp)$ around the on-shell condition,
$\Sigma(\rmp)=\Sigma_0(p)+\Sigma_1(p)(\rmp^2-m^2)+\mathcal{O}((\rmp^2-m^2)^2)$,
one obtains a spatial momentum dependent shift in the pole and a non-vanishing
residue $Z(p)$ such that
\begin{equation}
\begin{tikzpicture}[line width=1pt,baseline=-3]
\draw[-] (-1,0)--(0,0);
\draw [fill=black!20] (0.5,0) circle (0.5);
\draw[-] (1,0)--(2,0);
\end{tikzpicture}
=\frac{iZ(p)}{\rmp^2-m^2-\Sigma_0(p)} + \ldots\ .
\end{equation}
where $Z=1+h^{-1}\Sigma^{(1)}_1(p)+\mathcal{O}(h^{-2})$
and $\Sigma_0(p) = h^{-1}\Sigma_0^{(1)}(p) + \mathcal{O}(h^{-2})$.
It is well-known that
the same quantity also appears in the LSZ reduction and the prescription to
take these contributions into account is to include a factor of $\sqrt{Z}$ for
the external legs of the scattering process. When inserting this into the
S-matrix of a $2\to2$ process one therefore gets an additional contribution to
$T^{(1)}$ of the form
\begin{equation}\begin{split}\label{important}
T^{(1)}_{ext} = (\Sigma^{(1)}_1(p) + \Sigma^{(1)}_1(p'))T^{(0)}\ ,
\end{split}
\end{equation}
where we recall that we are working in the configuration in which $q =p$ and
$q'=p'$. Here we can already make the observation that given that
$\Sigma^{(1)}_1(p)$ is real (and assuming that $T^{(0)}$ is real) the
contribution from external legs should contribute to the real rational part of
$T^{(1)}$.

The contribution $\Sigma^{(1)}_1(p)$ is a subleading contribution in the
expansion of the self-energy around the on-shell condition and in a standard
Feynman diagram computation would be regularization dependent. Since in the
unitarity computation we did not assume any explicit regularization we may
encounter problems combining the two results. For this reason we will choose to
follow a rather different approach and compute this subleading contribution via
unitarity.

As we are considering a unitarity computation we will only consider
contributions from the graphs in figure \ref{graphs} when they have a physical
two-particle cut. In particular, to be consistent with our approach for the
S-matrix we ignore the latter two tadpole diagrams and restrict our attention
to the first diagram. It therefore follows that, in the unitarity computation,
external leg corrections will only play a role at one loop in theories with
cubic vertices. In the context of generalized unitarity, as we discussed in
section \ref{singlemass}, tadpole diagrams may not be negligible and therefore
there is no guarantee that our procedure will provide the whole result. However
the precise cancellation we observe in the specific example we discuss later is
a clear indication of the validity of our result up to a shift in the
coupling (for a more detailed discussion see section \ref{extlegs}).

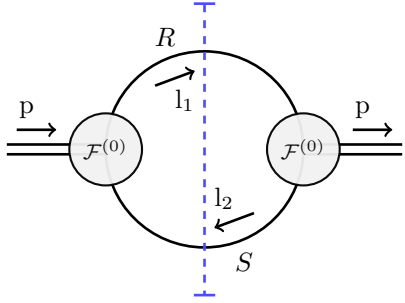
\begin{figure}[t]
\begin{center}
\begin{tikzpicture}[line width=1pt,scale=1.3,baseline = -100]
\draw[-] (-5,-4.05) -- (-4,-4.05);
\draw[-] (-5,-3.95) -- (-4,-3.95);
\draw[-] (-2,-4.05) -- (-1,-4.05);
\draw[-] (-2,-3.95) -- (-1,-3.95);
\draw (-3,-4) circle (1cm);
\draw[|-|,dashed,blue!70,line width=1pt] (-3,-2.5) -- (-3,-5.5);
\draw[->] (-4.9,-3.8) -- (-4.5,-3.8);
\node at (-4.8,-3.6) {$\rmp$};
\draw[->] (-1.5,-3.8) -- (-1.1,-3.8);
\node at (-1.4,-3.6) {$\rmp$};
\draw[<-] (-3.1,-3.2) -- (-3.5,-3.35);
\node at (-3.2,-3.5) {$\rml_1$};
\draw[<-] (-2.9,-4.8) -- (-2.5,-4.65);
\node at (-2.8,-4.5) {$\rml_2$};
\node at (-3.4,-2.85) {$\IndE$};
\node at (-2.6,-5.15) {$\IndF$};
\node [circle,draw=black!100,fill=black!5,thick,opacity=.95] at (-4,-4) {\footnotesize$\mathcal{F}^{(0)}$\normalsize};
\node [circle,draw=black!100,fill=black!5,thick,opacity=.95] at (-2,-4) {\footnotesize$\mathcal{F}^{(0)}$\normalsize};
\end{tikzpicture}
\caption{Cut of a two-point function obtained by fusing two form factors. The double line indicates an off-shell state.}
\label{2ptfun}\nonumber
\end{center}
\end{figure}

The computation of correlation functions by generalized unitarity was
extensively analyzed in four dimensions in \cite{Engelund:2012re} in which it
was shown that the object that needs to be put on either side of the cut is a
form factor\,\footnote{For the use of form factors in unitarity computations in
higher dimensions see also \cite{Brandhuber:2012vm}.\vspace{2pt}} as shown in figure
\ref{2ptfun}.
However, let us also note that we will want to expand around the on-shell
condition and hence we ask that the diagram should have a physical cut even
when the external leg is on-shell. This places a restriction on the masses of
the particles involved. In particular they should take the following form;
$m_1$, $m_2$ and $m_1 - m_2$, where we take $m_1>m_2$.

Now taking figure \ref{2ptfun} with a mass $m_1-m_2$ external
particle,\footnote{This will be the case we consider for $AdS_3 \times S^3
\times S^3 \times S^1$. One can also consider a mass $m_1$ external particle
and internal particles with masses $m_1-m_2$ and $m_2$ ($m_1>m_2$). In this
case the two loop momenta in figure \ref{2ptfun} should be pointing in the
same direction.\vspace{2pt}} internal particles with masses $m_1$ and $m_2$ corresponding
to momentum $\rml_1$ and $\rml_2$ and returning $\rmp$ off-shell, the explicit
expression for this diagram is given by
\begin{align}
\Sigma^{(1)}(\rmp)|_{cut}=\int\frac{d^2 \rml_1}{(2\pi)^2} \, i\pi\, \delta^+({\rml_1}^2-m_1)\, i\pi\, \delta^+((\rml_1-p)^2-m_2)\,
\mathcal{F}^{(0)}_{\IndE\IndF}(\rmp,\rml_1,\rml_1-\rmp)\,{\mathcal{F}^{(0)}_{\IndE\IndF}}^\dagger(\rmp,\rml_1,\rml_1-\rmp)\ .
\end{align}
Here, as in the unitarity computation of the S-matrix, the cut completely
freezes the internal momenta:
\begin{align}
\rml_1 &= \frac{m_1^2-m_2^2+\rmp^2-\sqrt{\Delta}}{2\, \rmp^2}\, \rmp\equiv \rml_*\ ,\\
\rmp-\rml_1 &= \frac{m_2^2-m_1^2+\rmp^2+\sqrt{\Delta}}{2\, \rmp^2}\, \rmp \equiv \rml'_*\ ,
\end{align}
where $\Delta=\rmp^4+m_1^4+m_2^4-2\, m_1^2 \rmp^2-2\, m_2^2 \rmp^2-2\, m_1^2
m_2^2$. It therefore follows that we can pull the numerators out of the
integrand and uplift the integral as was done for the four-point amplitude.
This gives
\begin{align}\label{ffsquare}
\Sigma^{(1)}(\rmp)=\frac12 \left|\mathcal{F}^{(0)}_{\IndE\IndF}(\rmp,\rml_*,\rml'_*)\right|^2 I(\rmp^2,m_1,m_2)\ ,
\end{align}
with the integral $I(\rmp^2,m_1,m_2)$ defined in \eqref{sbi}. In section
\ref{extlegs} we will apply this formula to a specific example and we will also
point out the limits of its application.

\subsection{Structure of the result}\label{resstruc}

To conclude this section let us make some remarks about the features of the
result that are relevant for the integrable light-cone gauge S-matrices for
strings in $AdS_3 \times S^3 \times M^4$ backgrounds. These theories all have
the property that the massive excitations can be grouped into particles and
antiparticles transforming with charge $\s = +1$ and $\s = -1$ under a global
$U(1)$ symmetry. Furthermore, not only is the set of incoming momenta preserved
by the scattering process, but so are the $U(1)$ charges associated to the
individual momenta, i.e. $\s_M = \s_P$ and $\s_N = \s_Q$. The general structure
of the S-matrix is then
\begin{align}
S_{MN}^{PQ}(p,p') &= \exp[i \vp_{\s_M \s_N}(p,p')] \hat{S}_{MN}^{PQ}(p,p') \label{structure}\ ,
\end{align}
where $\vp$ are the phases and the matrix structure $\hat{S}$ is fixed by the
symmetry of the theory. Each of these objects admit a perturbative expansion at
strong coupling (i.e. around $h = \infty$):
\begin{align}
S&=\mathbf{1}+i\sum_{n=1}^\infty h^{-n}T^{(n-1)} \ , \quad
\hat{S}=\mathbf{1}+i\sum_{n=1}^\infty h^{-n}\hat{T}^{(n-1)}\ ,
\quad \vp_{\s_M\s_N}(p,p')=\sum_{n=1}^\infty h^{-n}\vp^{(n-1)}_{\s_M\s_N}(p,p')\ . \label{phiexp1}
\end{align}
Furthermore, as $\hat{S}$ is fixed by symmetries it should contain no
logarithmic functions of the momenta. Therefore, all the logarithms are
contained in the phases, and to the one-loop order we can separate these off as
follows
\begin{equation}\label{phasestruc}
\vp^{(0)}_{\s_M\s_N}(p,p') = \phi^{(0)}_{\s_M\s_N}(p,p') \ , \qquad
\vp^{(1)}_{\s_M\s_N}(p,p') = \cl_{\s_M\s_N}(p,p')\,\theta + \phi^{(1)}_{\s_M\s_N}(p,p') \ .
\end{equation}
Here $\theta$, defined in eq.~\eqref{thetadef}, is the only possible logarithm
appearing at one loop, and $\phi^{(n)}_{\s_M\s_N}$ are rational functions of
the momenta.

Substituting eqs.~\eqref{phiexp1} and \eqref{phasestruc} into \eqref{structure}
we find
\begin{align}
T^{(0)}&= \phi^{(0)}_{\s_M\s_N}(p,p')\, \mathbf{1}+\hat{T}^{(0)}\ , \label{treestruc}\\
T^{(1)}&= \cl_{\s_M\s_N}(p,p')\, \theta\, \mathbf{1}+\frac i2\left[\phi^{(0)}_{\s_M\s_N}(p,p')\right]^2\mathbf{1}+\phi^{(1)}_{\s_M\s_N}(p,p')\, \mathbf{1}
+i\phi^{(0)}_{\s_M\s_N}(p,p')\,\hat{T}^{(0)}+\hat{T}^{(1)} \ . \label{1Lstruc}
\end{align}

Let us compare the structure of the one-loop result following from
integrability \eqref{1Lstruc} with that following from unitarity methods
\eqref{result}, \eqref{resulttwo}. The comparison between the two expressions
leads to the following identifications (note that by definition the functions
$\cl_{\s_M\s_N}$ and $\phi^{(n)}_{\s_M\s_N}$ are real)
\begin{align}
& \frac{1}{2\p} (T^{(0)} \cc{u} T^{(0)} - T^{(0)} \cc{s} T^{(0)})= \cl_{\s_M\s_N}(p,p')\,\mathbf{1}\ , \label{logsid}
\\
& \frac12 T^{(0)} \cc{s} T^{(0)}=\frac12\left[\phi^{(0)}_{\s_M\s_N}(p,p')\right]^2\mathbf{1}+\phi^{(0)}_{\s_M\s_N}(p,p')\,\hat{T}^{(0)}+\text{Im}(\hat{T}^{(1)}) \nonumber
\\ & \hspace{200pt} \Rightarrow \quad \frac12 \hat T^{(0)} \cc{s} \hat T^{(0)} = \text{Im}(\hat{T}^{(1)}) \ , \label{schannelid}\\
&
\frac1{16\p} (\frac{1}{m^2}\widetilde T^{(0)} \cct{t}{\la} T^{(0)}+\frac{1}{m'^2}T^{(0)} \cct{t}{\ra} \widetilde T^{(0)})
+ (\Sigma^{(1)}_1(p) + \Sigma^{(1)}_1(p')) T^{(0)} = \phi^{(1)}_{\s_M\s_N}(p,p')\, \mathbf{1}+ \text{Re}(\hat{T}^{(1)})\ ,\label{tchannelid}
\end{align}
where we have assumed that $T^{(0)}$ is real, which will indeed be the case for
all the models we consider. For the rational terms coming from the s-channel in
\eqref{schannelid} we have simplified the expression that needs to be checked
by substituting in for $T^{(0)}$ \eqref{treestruc} and using that $\mathbf{1}
\cc{s} \mathbf{1} = \mathbf{1}$, $\hat T^{(0)} \cc{s} \mathbf{1} = \hat T^{(0)}$
and $\mathbf{1} \cc{s} \hat T^{(0)} = \hat T^{(0)}$ are satisfied by definition
(see eq.~\eqref{scont}). In \eqref{tchannelid} we have also included a
possible contribution from external leg corrections to the real rational part
of $T^{(1)}$ (see eq.~\eqref{important}), as discussed in section
\ref{sec:elc}. Eqs.~\eqref{logsid}, \eqref{schannelid} and \eqref{tchannelid}
are therefore the three equations that we need to check to see how much of the
exact S-matrix is recovered from the unitarity construction.

Factoring out an overall phase factor as in \eqref{structure} clearly contains
a degree of arbitrariness. Of course, this choice should not affect the final
result, however, there are certain choices that interplay well with the
unitarity construction. In particular, if there is a scattering process for
which the only possible outgoing two-particle state is the incoming state
($M=P=M_*$, $N=Q=N_*$), then the corresponding amplitude must be a phase
factor. In this case we can set
\begin{equation}
\hat S_{M_* N_*}^{M_* N_*} = 1 \ ,
\end{equation}
where $M_*$ and $N_*$ are fixed and there is no sum. This choice is consistent
with \eqref{schannelid} -- both sides are clearly vanishing by construction.
Furthermore, $\phi^{(1)}$ is just given by the t-channel contraction (plus possible
external leg corrections) with indices $M=P=M_*$, $N=Q=N_*$.

\section{Massive sector for \txpf{$AdS_3 \times S^3 \times M^4$}{AdS3 x S3 x M4} supported RR flux}

In this section we apply the methods of section \ref{sec:gp} to the massive
sectors of the $AdS_3 \times S^3 \times T^4$ and $AdS_3 \times S^3 \times S^3
\times S^1$ light-cone gauge-fixed string theories supported by RR flux.

\subsection{Exact and tree-level S-matrices}\label{sec:exT4}

We start by reviewing some of the available results for the light-cone gauge
S-matrices for the $AdS_3 \times S^3 \times M^4$ string theories supported by
RR flux.

\subsubsection{Massive sector for \txpf{$AdS_3\times S^3 \times T^4$}{AdS3 x S3 x T4}}\label{sec:t4}

The quadratic light-cone gauge-fixed action for the $AdS_3
\times S^3 \times T^4$ background describes $4+4$ massive and $4+4$ massless
fields. Here we will just consider the scattering of two massive excitations to
two massive excitations. The S-matrix of the theory was fixed up to two phases
in \cite{Borsato:2013qpa} using symmetries.

Thinking of the particle content of the massive sector as $2+2$ complex degrees
of freedom, we label these fields as $\Phi_{\vf\vf}$, $\Phi_{\psi\psi}$,
$\Phi_{\vf\psi}$ and $\Phi_{\psi\vf}$, and their complex conjugates as
$\Phi_{\bar \vf\bar\vf}$, $\Phi_{\bar \psi \bar \psi}$, $\Phi_{\bar \vf\bar
\psi}$ and $\Phi_{\bar \psi\bar \vf}$, where we understand $\vf$, $\bar \vf$ as
bosonic and $\psi$, $\bar\psi$ as fermionic indices.

As a consequence of the symmetries and integrability of the theory, the
S-matrix factorizes:
\begin{equation}\label{factorize}
\mathbb{S}\ket{\Phi_{M\dot M}(p)\Phi_{N\dot N}(p')} = (-1)^{[\dot M][N]+[\dot P][Q]} S_{MN}^{PQ}(p,p')S_{\dot M \dot N}^{\dot P\dot Q}(p,p')
\ket{\Phi_{P\dot P}(p)\Phi_{Q\dot Q}(p')} \ ,
\end{equation}
where the indices take the following values:
$\{\vf,\bar{\vf},\psi,\bar{\psi}\}$. One can check that the construction outlined in section \ref{sec:gp} gives the same one-loop result whether we consider the factorized or full S-matrix. Therefore, for simplicity we will work with the former. The general structure of the factorized
S-matrix takes the form given in \eqref{structure} with $\s_\vf=\s_\psi=+$ and
$\s_{\bar{\vf}}=\s_{\bar{\psi}}=-$. Charge conjugation symmetry implies that
$\phi_{++}=\phi_{--}$, $\phi_{+-}=\phi_{-+}$, $\cl_{++}=\cl_{--}$ and
$\cl_{+-}=\cl_{-+}$. Therefore, in the following we will focus on the $++$ and
$+-$ sectors. A typical feature of the uniform light-cone gauge is the
dependence of the phase on a gauge-fixing parameter $a$. This dependence has
the following exact form
\begin{equation}\label{gauge}
\exp\big[\frac{i}{2}(a-\tfrac12)(\e'p-\e p')\big] \ ,
\end{equation}
where the all-order energies $\e$ are defined in appendix \ref{notations1}.

As we discussed in section \ref{resstruc} we define the overall phase factors
by setting particular components of $\hat{S}_{MN}^{PQ}$ to one
\begin{equation}
\hat{S}_{\vf\vf}^{\vf\vf}(p,p')=1\ , \qquad \hat{S}_{\vf\bar{\psi}}^{\vf\bar{\psi}}(p,p')=1\ . \label{choice1}
\end{equation}
The parametrizing functions of the exact S-matrix are defined as
\begin{align}
{S}_{\vf\vf}^{\vf\vf}(p,p')&=A_{++}(p,p') & {S}_{\vf\bar{\vf}}^{\vf\bar{\vf}}(p,p')&=A_{+-}(p,p')\nonumber\\
{S}_{\vf\psi}^{\vf\psi}(p,p')&=B_{++}(p,p') & {S}_{\vf\bar{\vf}}^{\psi\bar{\psi}}(p,p')&=B_{+-}(p,p')\nonumber\\
{S}_{\vf\psi}^{\psi\vf}(p,p')&=C_{++}(p,p') & {S}_{\vf\bar{\psi}}^{\vf\bar{\psi}}(p,p')&=C_{+-}(p,p')\nonumber\\
{S}_{\psi\vf}^{\psi\vf}(p,p')&=D_{++}(p,p') & {S}_{\psi\bar{\vf}}^{\psi\bar{\vf}}(p,p')&=D_{+-}(p,p')\nonumber\\
{S}_{\psi\vf}^{\vf\psi}(p,p')&=E_{++}(p,p') & {S}_{\psi\bar{\psi}}^{\psi\bar{\psi}}(p,p')&=E_{+-}(p,p')\nonumber\\
{S}_{\psi\psi}^{\psi\psi}(p,p')&=F_{++}(p,p') & {S}_{\psi\bar{\psi}}^{\vf\bar{\vf}}(p,p')&=F_{+-}(p,p')\label{corr}
\end{align}
with the functions in string frame given by \cite{Borsato:2013qpa}
\begin{align}
A_{++}(p,p') &= S^{11}_{++}(p,p') \ , &
B_{++}(p,p') &= S^{11}_{++}(p,p') \frac{x'^+ - x^+}{x'^+ - x^-} \frac{1}{\n}\ , \nonumber \\
C_{++}(p,p') &= S^{11}_{++}(p,p') \frac{x'^+ - x'^-}{x'^+ - x^-} \frac{\eta}{\eta'} \sqrt{\frac{\n'}{\n}}\ , &
D_{++}(p,p') &= S^{11}_{++}(p,p') \frac{x'^- - x^-}{x'^+ - x^-} \n'\ , \nonumber \\
E_{++}(p,p') &= S^{11}_{++}(p,p') \frac{x^+ - x^-}{x'^+ - x^-} \frac{\eta'}{\eta} \sqrt{\frac{\n'}{\n}}\ , &
F_{++}(p,p') &= S^{11}_{++}(p,p') \frac{x'^- - x^+}{x'^+ - x^-}\frac{\n'}{\n}\ , \label{funpp}
\end{align}
\begin{align}
A_{+-}(p,p') &= S^{11}_{+-}(p,p') \frac{1-\frac{1}{x^+\, x'^-}}{1-\frac{1}{x^-\, x'^-}} \n\ , &
B_{+-}(p,p') &= -S^{11}_{+-}(p,p') \frac{i\, \eta\eta'}{x^- x'^-}\frac{1}{1-\frac{1}{x^-\, x'^-}} (\n\n')^{-\frac12}\ , \nonumber \\
C_{+-}(p,p') &= S^{11}_{+-}(p,p')\ , &
D_{+-}(p,p') &= S^{11}_{+-}(p,p') \frac{1-\frac{1}{x^+\, x'^+}}{1-\frac{1}{x^-\, x'^-}}\n\n'\ , \nonumber \\
E_{+-}(p,p') &= S^{11}_{+-}(p,p') \frac{1-\frac{1}{x^-\, x'^+}}{1-\frac{1}{x^-\, x'^-}} \n'\ , &
F_{+-}(p,p') &= -S^{11}_{+-}(p,p') \frac{i\, \eta\eta'}{x^+ x'^+}\frac{1}{1-\frac{1}{x^-\, x'^-}}(\n\n')^\frac32\ . \label{funpm}
\end{align}
The definitions of the variables $x^\pm$ entering these expressions are given
for general mass in appendix \ref{notations1}. Here the masses should be set to
one. The functions $S^{11}_{++}(p,p')$ and $S^{11}_{+-}(p,p')$ are two overall
phase factors, i.e. in the notation of eq.~\eqref{phasestruc}
$S^{11}_{\s_M\s_N}(p,p')=e^{i\vp^{11}_{\s_M\s_N}(p,p')}$. The superscripts
refer to the masses of the two particles being scattered. These phase factors
are not fixed by symmetry. They are, however, constrained by crossing symmetry
and a conjecture for their exact expressions was given in
\cite{Borsato:2013hoa}, supported by semiclassical one-loop computations
\cite{Abbott:2012dd,Beccaria:2012kb} (see footnote \ref{clarification}). 
More details are given in appendix \ref{phases}.

The input needed to apply the unitarity construction described in section
\ref{sec:gp} is the tree-level S-matrix. Various components were computed
directly in \cite{Sundin:2013ypa,Rughoonauth:2012qd}. These are consistent with
the near-BMN expansion of the exact result \eqref{funpp}, \eqref{funpm}, for
which we recall that the integrable coupling used in the definition of $x^\pm$
has the expansion $\hh(h) = h+\mathcal{O}(h^0)$ and to take the near-BMN limit
the spatial momenta should first be rescaled; $p\to \frac{p}{h}$. The remaining
components of the tree-level S-matrix can then be fixed from the expansion of
the exact result. Here we shall present the result in the gauge $a=\frac12$ as
the dependence on $a$ goes through the unitarity procedure without any
particular subtlety, i.e. it exponentiates as in eq.~\eqref{gauge}. The
tree-level S-matrix reads
\begin{align}
A^{(0)}_{++}(p,p') &= \frac{(p+p')^2}{4(e' p-e p')}\ ,
& B^{(0)}_{++}(p,p') &= \frac{p'^2-p^2}{4(e' p-e p')}\ , \nonumber \\
C^{(0)}_{++}(p,p') &= p p' \frac{\sqrt{(e+p)(e'-p')}+\sqrt{(e-p)(e'+p')}}{2(e' p-e p')}\ ,
& D^{(0)}_{++}(p,p') &= -\frac{p'^2-p^2}{4(e' p-e p')}\ , \nonumber \\
E^{(0)}_{++}(p,p') &= p p' \frac{\sqrt{(e+p)(e'-p')}+\sqrt{(e-p)(e'+p')}}{2(e' p-e p')}\ ,
& F^{(0)}_{++}(p,p') &= -\frac{(p+p')^2}{4(e' p-e p')}\ , \label{treepp}
\end{align}
\begin{align}
A^{(0)}_{+-}(p,p') &= \frac{(p-p')^2}{4(e' p-e p')}\ ,
& B^{(0)}_{+-}(p,p') &= p p' \frac{\sqrt{(e-p)(e'+p')}-\sqrt{(e+p)(e'-p')}}{2(e' p-e p')}\ , \nonumber \\
C^{(0)}_{+-}(p,p') &= \frac{p'^2-p^2}{4(e' p-e p')}\ ,
& D^{(0)}_{+-}(p,p') &= -\frac{p'^2-p^2}{4(e' p-e p')}\ , \nonumber \\
E^{(0)}_{+-}(p,p') &= -\frac{(p-p')^2}{4(e' p-e p')}\ ,
& F^{(0)}_{+-}(p,p') &= p p' \frac{\sqrt{(e-p)(e'+p')}-\sqrt{(e+p)(e'-p')}}{2(e' p-e p')}\ . \label{treepm}
\end{align}

\subsubsection{Massive sector for \txpf{$AdS_3\times S^3 \times S^3 \times S^1$}{AdS3 x S3 x S3 x S1}}

The quadratic light-cone gauge-fixed action for the $AdS_3 \times S^3
\times S^3 \times S^1$ background describes particles with four different
masses. The field content is summarised in table \ref{masses}. Here we will
focus on the scattering of massive states with masses $\a$ and $\bar\a=1-\a$.
\begin{table}[ht]
\begin{center}
\begin{tabular}[h]{|c|l|}
\hline
Fields& Mass\\
\hline
$\varphi_1,\bar{\varphi}_1,\chi^1,\bar{\chi}^1$&$m_1=1$\\
\hline
$\varphi_2,\bar{\varphi}_2,\chi^2,\bar{\chi}^2$&$m_2=\a$\\
\hline
$\varphi_3,\bar{\varphi}_3,\chi^3,\bar{\chi}^3$&$m_3=\bar\a$\\
\hline
$\varphi_4,\bar{\varphi}_4,\chi^4,\bar{\chi}^4$&$m_4=0$\\
\hline
\end{tabular}
\end{center}
\caption{Field content of the $AdS_3\times S^3 \times S^3 \times S^1$ light-cone gauge-fixed string theory.}\label{masses}
\end{table}

Let us first analyze the S-matrix for $AdS_3\times S^3 \times S^3 \times S^1$
describing the scattering of two particles of mass $\a$.\footnote{For particles
of mass $\bar\a$ the corresponding result can be obtained simply by replacing
$\a$ with $\bar\a$.\vspace{2pt}} When we restrict to this sector the S-matrix has the same
structure as the factorized S-matrix for $AdS_3\times S^3 \times T^4$, again
taking the form given in \eqref{structure}. The tree-level S-matrix, however,
is different and this will have non-trivial consequences for the unitarity
calculation. Compared to the $AdS_3\times S^3 \times T^4$ case the dependence
on the gauge-fixing parameter $a$ is modified due to the fact that this is now
the full S-matrix. The new expression reads
\begin{equation}\label{gaugeS1}
\exp\big[i(a-\tfrac12)(\e'p-\e p')\big] \ .
\end{equation}

We again use \eqref{choice1} to choose the overall phase
factors and define the parametrizing functions as in
eq.~\eqref{corr}.\footnote{To be precise we use the definitions \eqref{corr}
with the replacements $\varphi\to \varphi_2$ and $\psi\to\chi^2$ and likewise
for their conjugates.\vspace{2pt}} The exact result reads
\cite{Borsato:2012ud,Borsato:2012ss}
\begin{align}
A_{++}(p,p') &= S^{\a\a}_{++}(p,p') \ , &
B_{++}(p,p') &= S^{\a\a}_{++}(p,p') \frac{x'^+ - x^+}{x'^+ - x^-} \frac{1}{\n}\ , \nonumber \\
C_{++}(p,p') &= S^{\a\a}_{++}(p,p') \frac{x'^+ - x'^-}{x'^+ - x^-} \frac{\eta}{\eta'} \sqrt{\frac{\n'}{\n}}\ , &
D_{++}(p,p') &= S^{\a\a}_{++}(p,p') \frac{x'^- - x^-}{x'^+ - x^-} \n'\ , \nonumber \\
E_{++}(p,p') &= S^{\a\a}_{++}(p,p') \frac{x^+ - x^-}{x'^+ - x^-} \frac{\eta'}{\eta} \sqrt{\frac{\n'}{\n}}\ , &
F_{++}(p,p') &= S^{\a\a}_{++}(p,p') \frac{x'^- - x^+}{x'^+ - x^-}\frac{\n'}{\n}\ , \label{funppS1}
\end{align}
\begin{align}
A_{+-}(p,p') &= S^{\a\a}_{+-}(p,p') \frac{1-\frac{1}{x^+\, x'^-}}{1-\frac{1}{x^-\, x'^-}} \n\ , &
B_{+-}(p,p') &= -S^{\a\a}_{+-}(p,p') \frac{i\, \eta\eta'}{x^- x'^-}\frac{1}{1-\frac{1}{x^-\, x'^-}} (\n\n')^{-\frac12}\ , \nonumber \\
C_{+-}(p,p') &= S^{\a\a}_{+-}(p,p')\ , &
D_{+-}(p,p') &= S^{\a\a}_{+-}(p,p') \frac{1-\frac{1}{x^+\, x'^+}}{1-\frac{1}{x^-\, x'^-}}\n\n'\ , \nonumber \\
E_{+-}(p,p') &= S^{\a\a}_{+-}(p,p') \frac{1-\frac{1}{x^-\, x'^+}}{1-\frac{1}{x^-\, x'^-}} \n'\ , &
F_{+-}(p,p') &= -S^{\a\a}_{+-}(p,p') \frac{i\, \eta\eta'}{x^+ x'^+}\frac{1}{1-\frac{1}{x^-\, x'^-}}(\n\n')^\frac32\ . \label{funpmS1}
\end{align}
The structure of the S-matrix is identical to \eqref{funpp} and \eqref{funpm},
the only differences being the overall phase factors, $S^{\a\a}_{++}(p,p')$ and
$S^{\a\a}_{+-}(p,p')$, and that in the definition of the variables $x^\pm$
given in appendix \ref{notations1} the mass should be set to $\a$. The phase
factors $S^{\a\a}_{\pm\pm}$ and $S^{\a\a}_{\pm\mp}$ have been computed
semiclassically in \cite{Abbott:2013ixa}.

Various components of the tree-level S-matrix were computed directly in
\cite{Sundin:2013ypa,Rughoonauth:2012qd}. These are consistent with the
near-BMN expansion of the exact result \eqref{funppS1}, \eqref{funpmS1}. As in
the $AdS_3 \times S^3 \times T^4$ case, the remaining components of the
tree-level S-matrix can then be fixed from the expansion of the exact result.
We shall present the result in the gauge $a=\frac12$ as the dependence on $a$
again goes through the unitarity procedure without any particular subtlety,
i.e. it exponentiates as in eq.~\eqref{gaugeS1}. The tree-level S-matrix reads
\begin{align}
A^{(0)}_{++}(p,p') &= \frac{\a (p+p')^2}{2(e' p-e p')}\ ,
& B^{(0)}_{++}(p,p') &= \frac{\a p' (p+p')}{2(e' p-e p')}\ , \nonumber \\
C^{(0)}_{++}(p,p') &= p p' \frac{\sqrt{(e+p)(e'-p')}+\sqrt{(e-p)(e'+p')}}{2(e' p-e p')}\ , &
D^{(0)}_{++}(p,p') &= \frac{\a p (p+p')}{2(e' p-e p')}\ , \nonumber \\
E^{(0)}_{++}(p,p') &= p p' \frac{\sqrt{(e+p)(e'-p')}+\sqrt{(e-p)(e'+p')}}{2(e' p-e p')}\ , &
F^{(0)}_{++}(p,p') &= 0\ ,\label{tree3}
\end{align}
\begin{align}
A^{(0)}_{+-}(p,p') &= \frac{\a(p-p')^2}{2(e' p-e p')}\ , &
B^{(0)}_{+-}(p,p') &= p p' \frac{\sqrt{(e-p)(e'+p')}-\sqrt{(e+p)(e'-p')}}{2(e' p-e p')}\ , \nonumber \\
C^{(0)}_{+-}(p,p') &= \frac{\a p' (p'-p)}{2(e' p-e p')}\ , &
D^{(0)}_{+-}(p,p') &= \frac{\a p (p-p')}{2(e' p-e p')}\ , \nonumber \\
E^{(0)}_{+-}(p,p') &= 0\ , &
F^{(0)}_{+-}(p,p') &= p p' \frac{\sqrt{(e-p)(e'+p')}-\sqrt{(e+p)(e'-p')}}{2(e' p-e p')}\ .\label{tree4}
\end{align}

Let us now turn our attention to the scattering between a mode with mass $\a$
and one with mass $\bar\a = 1-\a$. There are no surprises regarding the
gauge-fixing parameter $a$, i.e. eq.~\eqref{gaugeS1} also holds for the two
mass scattering. We again define the parametrizing functions as
\begin{align}
{S}_{\vf_2\vf_3}^{\vf_2\vf_3}(p,p')&=A_{++}(p,p') & {S}_{\vf_2\bar{\vf_3}}^{\vf_2\bar{\vf_3}}(p,p')&=A_{+-}(p,p')\nonumber\\
{S}_{\vf_2\chi^3}^{\vf_2\chi^3}(p,p')&=B_{++}(p,p') & {S}_{\vf_2\bar{\vf_3}}^{\chi^2\bar{\chi}^3}(p,p')&=B_{+-}(p,p')\nonumber\\
{S}_{\vf_2\chi^3}^{\chi^2\vf_3}(p,p')&=C_{++}(p,p') & {S}_{\vf_2\bar{\chi}^3}^{\vf_2\bar{\chi}^3}(p,p')&=C_{+-}(p,p')\nonumber\\
{S}_{\chi^2\vf_3}^{\chi^2\vf_3}(p,p')&=D_{++}(p,p') & {S}_{\chi^2\bar{\vf_3}}^{\chi^2\bar{\vf_3}}(p,p')&=D_{+-}(p,p')\nonumber\\
{S}_{\chi^2\vf_3}^{\vf_2\chi^3}(p,p')&=E_{++}(p,p') & {S}_{\chi^2\bar{\chi}^3}^{\chi^2\bar{\chi}^3}(p,p')&=E_{+-}(p,p')\nonumber\\
{S}_{\chi^2\chi^3}^{\chi^2\chi^3}(p,p')&=F_{++}(p,p') & {S}_{\chi^2\bar{\chi}^3}^{\vf_2\bar{\vf_3}}(p,p')&=F_{+-}(p,p')
\end{align}
with the functions in string frame given by \cite{Borsato:2012ud,Borsato:2012ss}
\begin{align}
A_{++}(p,p') &= S^{\a\bar \a}_{++}(p,p') \ , &
B_{++}(p,p') &= S^{\a\bar \a}_{++}(p,p') \frac{y'^+ - x^+}{y'^+ - x^-} \frac{1}{\n}\ , \nonumber \\
C_{++}(p,p') &= S^{\a\bar \a}_{++}(p,p') \frac{y'^+ - y'^-}{y'^+ - x^-} \frac{\eta}{\eta'}\sqrt{\frac{\n'}{\n}}\ , &
D_{++}(p,p') &= S^{\a\bar \a}_{++}(p,p') \frac{y'^- - x^-}{y'^+ - x^-} \n'\ , \nonumber \\
E_{++}(p,p') &= S^{\a\bar \a}_{++}(p,p') \frac{x^+ - x^-}{y'^+ - x^-} \frac{\eta'}{\eta} \sqrt{\frac{\n'}{\n}}\ , &
F_{++}(p,p') &= S^{\a\bar \a}_{++}(p,p') \frac{y'^- - x^+}{y'^+ - x^-} \frac{\n'}{\n}\ , \label{funpppS1}
\end{align}
\begin{align}
A_{+-}(p,p') &= S^{\a\bar \a}_{+-}(p,p') \frac{1-\frac{1}{x^+\, y'^-}}{1-\frac{1}{x^-\, y'^-}} \n\ , &
B_{+-}(p,p') &= -S^{\a\bar \a}_{+-}(p,p') \frac{i\, \eta\eta'}{x^-y'^-}\frac{1}{1-\frac{1}{x^-\, y'^-}} (\n\n')^{-\frac12}\ , \nonumber \\
C_{+-}(p,p') &= S^{\a\bar \a}_{+-}(p,p') \ , &
D_{+-}(p,p') &= S^{\a\bar \a}_{+-}(p,p') \frac{1-\frac{1}{x^+\, y'^+}}{1-\frac{1}{x^-\, y'^-}}\n\n'\ , \nonumber \\
E_{+-}(p,p') &= S^{\a\bar \a}_{+-}(p,p') \frac{1-\frac{1}{x^-\, y'^+}}{1-\frac{1}{x^-\, y'^-}} \n'\ , &
F_{+-}(p,p') &= -S^{\a\bar \a}_{+-}(p,p') \frac{i\, \eta\eta'}{x^+ y'^+}\frac{1}{1-\frac{1}{x^-\, y'^-}}(\n\n')^\frac32\ . \label{funppmS1}
\end{align}
Here we have defined the overall phase factors by setting
\begin{equation}
{\hat S}_{\vf_2\vf_3}^{\vf_2\vf_3}(p,p')=1\ , \qquad {\hat S}_{\vf_2\bar{\chi}^3}^{\vf_2\bar{\chi}^3}(p,p')=1\ .
\end{equation}
The phase factors $S^{\a\bar\a}_{\pm\pm}$ and $S^{\a\bar\a}_{\pm\mp}$ have been computed semiclassically at one loop in \cite{Abbott:2013ixa}.

As before, the tree-level S-matrix can be extracted from the near-BMN expansion
of the exact result along with those amplitudes computed in
\cite{Sundin:2013ypa,Rughoonauth:2012qd}. For $a=\frac12$ (again the
contribution of the gauge-fixing parameter $a$ to the unitarity computation
goes through without any particular subtlety) it is given by
\begin{align}
{A}^{(0)}_{++}(p,p') &= 0\ , &
{B}^{(0)}_{++}(p,p') &=-\frac{ p (\bar\a p+\a p')}{2(e' p-e p')}\ , \nonumber \\
{C}^{(0)}_{++}(p,p') &= p p' \frac{\sqrt{(e+p)(e'-p')}+\sqrt{(e-p)(e'+p')}}{2(e' p-e p')}\ , &
{D}^{(0)}_{++}(p,p') &= -\frac{ p' (\bar\a p+\a p')}{2(e' p-e p')}\ , \nonumber \\
{E}^{(0)}_{++}(p,p') &= p p' \frac{\sqrt{(e+p)(e'-p')}+\sqrt{(e-p)(e'+p')}}{2(e' p-e p')}\ , &
{F}^{(0)}_{++}(p,p') &= -\frac{(p+p')(\bar\a p+\a p')}{2(e' p-e p')}\ ,\label{tree5}
\end{align}
\begin{align}
{A}^{(0)}_{+-}(p,p') &= 0\ , &
{B}^{(0)}_{+-}(p,p') &= p p' \frac{\sqrt{(e-p)(e'+p')}-\sqrt{(e+p)(e'-p')}}{2(e' p-e p')}\ , \nonumber \\
{C}^{(0)}_{+-}(p,p') &= -\frac{ p (\bar\a p-\a p')}{2(e' p-e p')} \ , &
{D}^{(0)}_{+-}(p,p') &= \frac{ p' (\bar\a p-\a p')}{2(e' p-e p')}\ , \nonumber \\
{E}^{(0)}_{+-}(p,p') &= -\frac{(p-p')(\bar\a p-\a p')}{2(e' p-e p')}\ , &
{F}^{(0)}_{+-}(p,p') &= p p' \frac{\sqrt{(e-p)(e'+p')}-\sqrt{(e+p)(e'-p')}}{2(e' p-e p')}\ .\label{tree6}
\end{align}

\subsubsection{A general tree-level S-matrix for the \txpf{$AdS_3\times S^3 \times M^4$}{AdS3 x S3 x M4} theories}\label{general}

Comparing the expressions \eqref{treepp}, \eqref{treepm}, \eqref{tree3},
\eqref{tree4}, \eqref{tree5} and \eqref{tree6} we notice their similarity. In
particular, they all differ from one another by a term proportional to the
identity. Therefore in this section we will introduce an additional parameter
$\b$ along with two generic masses $m$ and $m'$, such that, for particular
values of these three parameters the tree-level S-matrices are recovered. The
advantage of this approach is that it demonstrates how some quantities in the
one-loop result are common to all three theories (i.e. $\b$-independent) up to
the right assignment of the masses.

To be concrete the expression for the general tree-level S-matrix is (we use the
notation $\bar \beta=(1-\beta)$)
\begin{align}
{A}^{(0)}_{++}(p,p') &= \beta \frac{(p+p')(m' p+m p')}{2(e' p-e p')}\ , &
{B}^{(0)}_{++}(p,p') &=\frac{ ( \beta p'-\bar \b p) (m' p+m p')}{2(e' p-e p')}\ , \nonumber \\
{C}^{(0)}_{++}(p,p') &= p p' \frac{\sqrt{(e+p)(e'-p')}+\sqrt{(e-p)(e'+p')}}{2(e' p-e p')}\ , &
{D}^{(0)}_{++}(p,p') &= \frac{ (\beta p-\bar\beta p') (m' p+m p')}{2(e' p-e p')}\ , \nonumber \\
{E}^{(0)}_{++}(p,p') &= p p' \frac{\sqrt{(e+p)(e'-p')}+\sqrt{(e-p)(e'+p')}}{2(e' p-e p')}\ , &
{F}^{(0)}_{++}(p,p') &= -\bar\beta \frac{(p+p')(m' p+m p')}{2(e' p-e p')}\ ,\nonumber
\end{align}
\begin{align}
{A}^{(0)}_{+-}(p,p') &= \b \frac{(p-p')(m' p-m p')}{2(e' p-e p')} \ , &
{B}^{(0)}_{+-}(p,p') &= p p' \frac{\sqrt{(e-p)(e'+p')}-\sqrt{(e+p)(e'-p')}}{2(e' p-e p')}\ , \nonumber \\
{C}^{(0)}_{+-}(p,p') &= \frac{ (\bar\b p+\b p') (m p'-m' p)}{2(e' p-e p')} \ , &
{D}^{(0)}_{+-}(p,p') &= \frac{ (\bar\b p'+\b p) (m' p-m p')}{2(e' p-e p')}\ , \nonumber \\
{E}^{(0)}_{+-}(p,p') &= -\bar\b \frac{(p-p')(m' p-m p')}{2(e' p-e p')}\ , &
{F}^{(0)}_{+-}(p,p') &= p p' \frac{\sqrt{(e-p)(e'+p')}-\sqrt{(e+p)(e'-p')}}{2(e' p-e p')}\ . \label{interp}
\end{align}
The explicit assignments that need to be made to recover the various tree-level
S-matrices given in the previous section are shown in table \ref{assign}. For
most of the unitarity computation however, we will keep general values of
$\beta$, $m$ and $m'$ so as to better understand the dependence of the result
on these parameters.
\begin{table}[ht]
\begin{center}
\begin{tabular}[h]{|c|c|}
\hline
$(\b,m,m')$&\textbf{Theory}\\
\hline
$(0,\a,\bar\a)$&$AdS_3\times S^3\times S^3\times S^1$ (two mass scattering)\\
\hline
$(\frac12,1,1)$&$AdS_3\times S^3\times T^4$\\
\hline
$(1,\a,\a)$&$AdS_3\times S^3\times S^3\times S^1$ (one mass scattering)\\
\hline
\end{tabular}
\end{center}
\caption{Assignments of parameters for the various theories of interest.}\label{assign}
\end{table}

\subsection{Result from unitarity techniques}

In this section we compute the one-loop S-matrix from unitarity methods for the
light-cone gauge-fixed string theories in the $AdS_3 \times S^3 \times T^4$ and
$AdS_3 \times S^3 \times S^3 \times S^1$ backgrounds supported by RR flux. As
explained in section \ref{resstruc}, we will split the result according to
eqs.~\eqref{logsid}, \eqref{schannelid} and \eqref{tchannelid}, where we recall
that we have chosen $S^{\vf\vf}_{\vf\vf}=A_{++}(p,p')$ and $S^{\vf
\bar\psi}_{\vf \bar\psi}=C_{+-}(p,p')$ as the overall phase factors.

In the general construction described in section \ref{2mass}, we found that
when scattering a particle of mass $m$ with one of mass $m'$, the s- and
u-channel contributions are just
given in terms of the tree-level S-matrices for the same scattering
configuration. Therefore, as the logarithmic terms \eqref{logsid} and the
rational terms \eqref{schannelid} only come from the s-channel
and u-channel contributions, for these we can work with the general
($\b$-dependent) tree-level S-matrix \eqref{interp}. For the t-channel
contribution \eqref{tchannelid} one needs to combine different tree-level
S-matrices, for example the scattering of two particles of mass $m$ with the
scattering of a particle of mass $m$ with one of mass $m'$. Hence for these
terms we will need to restrict to the specific values of $\b$, $m$ and $m'$
given in table \ref{assign}.

\subsubsection{Coefficients of the logarithms}

We start by briefly reviewing the work of \cite{Engelund:2013fja}. As discussed
in section \ref{resstruc} one should always be able to include the logarithmic
terms of the S-matrix in the phases. Therefore at one loop we expect them to
only contribute to the diagonal terms. In \cite{Engelund:2013fja} the authors
proved that this is indeed the case and that furthermore, the particular
combination governing the logarithmic dependence does not depend on the diagonal
components of the tree-level S-matrix. Therefore, the one-loop logarithmic
terms following from the unitarity construction for the general tree-level
S-matrix \eqref{interp} will be $\b$-independent. Indeed,
\begin{align}
\cl_{++}(p,p')&=-\frac{p^2p'^2}{4 \p (ee'-pp'-mm')}\ , \label{lppS1}\\
\cl_{+-}(p,p')&=-\frac{p^2p'^2}{4 \p (ee'-pp'+mm')}\ , \label{lpmS1}
\end{align}
where the functions $\cl_{\s_M\s_N}$ were introduced in eq.~\eqref{phasestruc}.
Although not transparent from this expression, these functions can be expressed
as
\begin{align}
\cl_{++}(p,p')&=-\frac{1}{2\pi}{C}^{(0)}_{++}(p,p'){E}^{(0)}_{++}(p,p')\ ,\label{logs1}\\
\cl_{+-}(p,p')&=-\frac{1}{2\pi}{B}^{(0)}_{+-}(p,p'){F}^{(0)}_{+-}(p,p')\ .\label{logs2}
\end{align}
As pointed out in \cite{Engelund:2013fja} the fact that these functions can be
expressed in terms of the entries of the tree-level S-matrix is clear from the
unitarity construction. In the next section we will show how this property
extends to the rational part following from the s-channel.

\subsubsection{Rational terms from the s-channel}

In section \ref{resstruc} we described how the contributions to the rational
part of the S-matrix in the unitarity calculation are split between the
s-channel \eqref{schannelid} and t-channel \eqref{tchannelid}. Let us start by
considering the s-channel, for which we can work with the general
$\b$-dependent tree-level S-matrix \eqref{interp}. From eq.~\eqref{schannelid}
it is clear that we can restrict our attention to $\text{Im}(\hat{T}^{(1)})$,
where we recall that $\hat{T}^{(0)}$ and $\hat{T}^{(1)}$ are the tree-level and
one-loop terms in the expansion of the S-matrix with the overall phase factors,
$S^{\vf\vf}_{\vf\vf}=A_{++}(p,p')$ and $S^{\vf \bar\psi}_{\vf\bar\psi}=C_{+-}(p,p')$,
set to one. The result from the unitarity calculation
is \eqref{schannelid}
\begin{align}
\frac12 \hat{T}^{(0)} \cc{s} \hat{T}^{(0)} \ . \label{subtraction}
\end{align}
Below we give the components of \eqref{subtraction}. These are in perfect
agreement with the one-loop expansion of the exact results \eqref{funpp},
\eqref{funpm}, \eqref{funppS1}, \eqref{funpmS1}, \eqref{funpppS1} and
\eqref{funppmS1} for the appropriate assignments of the masses $m$ and $m'$,
see table \ref{assign}. The one-loop expressions are
\begin{align}
{\hat A}^{(1)}_{++}(p,p') &= 0\ , \nonumber \\
{\hat B}^{(1)}_{++}(p,p') &=\frac12\Big[\frac{p(m'p+mp')}{2 (e'p-ep')}\Big]^2+\frac12 \Big[p p' \frac{\sqrt{(e+p)(e'-p')}+\sqrt{(e-p)(e'+p')}}{2(e' p-e p')}\Big]^2\ , \nonumber \\
{\hat{C}}^{(1)}_{++}(p,p') &=-\frac12\Big[\frac{(p+p')(m'p+mp')}{2(e'p-ep')}\Big]\Big[p p' \frac{\sqrt{(e+p)(e'-p')}+\sqrt{(e-p)(e'+p')}}{2(e' p-e p')}\Big]\ , \nonumber \\
{\hat D}^{(1)}_{++}(p,p') &= \frac12\Big[\frac{p'(m'p+mp')}{2 (e'p-ep')}\Big]^2+\frac12 \Big[p p' \frac{\sqrt{(e+p)(e'-p')}+\sqrt{(e-p)(e'+p')}}{2(e' p-e p')}\Big]^2\ , \nonumber \\
{\hat E}^{(1)}_{++}(p,p') &= -\frac12\Big[\frac{(p+p')(m'p+mp')}{2(e'p-ep')}\Big]\Big[p p' \frac{\sqrt{(e+p)(e'-p')}+\sqrt{(e-p)(e'+p')}}{2(e' p-e p')}\Big]\ , \nonumber \\
{\hat F}^{(1)}_{++}(p,p') &= \frac12\Big[\frac{(p+p')(m'p+mp')}{2(e'p-ep')}\Big]^2\ . \label{onelooppp}\\
{\hat A}^{(1)}_{+-}(p,p')&=\frac12\Big[\frac{p(m'p-mp')}{2 (e'p-ep')}\Big]^2+\frac12 \Big[p p' \frac{\sqrt{(e+p)(e'-p')}-\sqrt{(e-p)(e'+p')}}{2(e' p-e p')}\Big]^2\ , \nonumber \\
{\hat B}^{(1)}_{+-}(p,p')&=-\frac12\Big[\frac{(p+p')(m'p-mp')}{2(e'p-ep')}\Big]\Big[p p' \frac{\sqrt{(e+p)(e'-p')}-\sqrt{(e-p)(e'+p')}}{2(e' p-e p')}\Big]\ ,\nonumber \\
{\hat C}^{(1)}_{+-}(p,p')&=0\ ,\nonumber \\
{\hat D}^{(1)}_{+-}(p,p')&=\frac12\Big[\frac{(p+p')(m'p-mp')}{2(e'p-ep')}\Big]^2\ ,\nonumber \\
{\hat E}^{(1)}_{+-}(p,p')&=\frac12\Big[\frac{p'(m'p-mp')}{2 (e'p-ep')}\Big]^2+\frac12 \Big[p p' \frac{\sqrt{(e+p)(e'-p')}-\sqrt{(e-p)(e'+p')}}{2(e' p-e p')}\Big]^2\ ,\nonumber \\
{\hat F}^{(1)}_{+-}(p,p')&=-\frac12\Big[\frac{(p+p')(m'p-mp')}{2(e'p-ep')}\Big]\Big[p p' \frac{\sqrt{(e+p)(e'-p')}-\sqrt{(e-p)(e'+p')}}{2(e' p-e p')}\Big]\ .\label{onelooppm}
\end{align}
Although there are simpler ways to express this result, we have chosen this
form in order to explicitly show the connection with the tree-level functions.
The $\b$-independence of \eqref{onelooppp} and \eqref{onelooppm} is expected
since $\b$ appears only in the phases. As explained earlier in this section
and in section \ref{resstruc}, to check the s-channel rational terms we do not need
to consider the overall phase factors and hence they have been set to one.

Note that expressions for the components of $\frac12 T^{(0)} \cc{s}
T^{(0)}$ in terms of tree-level functions are given in \cite{Engelund:2013fja}
for $AdS_3\times S^3\times T^4$. These formulae also hold for the general
tree-level S-matrix \eqref{interp}, however, they will depend on $\b$, which
drops out only if we consider $\frac12 \hat{T}^{(0)} \cc{s} \hat{T}^{(0)}$ as
above.  To see explicitly how this works let us consider $F_{++}$.\footnote{For
the remainder of this section the dependence on $p$ and $p'$ is understood.\vspace{2pt}}
From \cite{Engelund:2013fja} the one-loop expression for $F_{++}$ is simply
given by
\begin{equation}
F^{(1)}_{++}=\frac12 [F^{(0)}_{++}]^2\ ,
\end{equation}
however when we consider \eqref{subtraction} (taking into account that
$\phi^{(0)}_{++}=A^{(0)}_{++}$) we find
\begin{equation}
\hat F^{(1)}_{++}=\frac12 [\hat F^{(0)}_{++}]^2=\frac12 [F^{(0)}_{++}-A_{++}^{(0)}]^2\ .
\end{equation}
Comparing the expressions for $F^{(0)}_{++}$ and $A_{++}^{(0)}$ we can then
observe the cancellation of $\b$. A similar story holds for the other
components
\begin{align}
{\hat A}^{(1)}_{++} &= 0 \ ,&
{\hat B}^{(1)}_{++} &=\frac12 [B^{(0)}_{++}-A_{++}^{(0)}]^2+\frac12 {C}^{(0)}_{++} {E}^{(0)}_{++} \ , \nonumber \\
{\hat{C}}^{(1)}_{++} &=\frac12 [B^{(0)}_{++}+D_{++}^{(0)}-2 A_{++}^{(0)}]\,{C}^{(0)}_{++}\ , &
{\hat D}^{(1)}_{++} &= \frac12 [D^{(0)}_{++}-A_{++}^{(0)}]^2+\frac12 {C}^{(0)}_{++} {E}^{(0)}_{++}\ , \nonumber \\
{\hat E}^{(1)}_{++} &= \frac12 [B^{(0)}_{++}+D_{++}^{(0)}-2 A_{++}^{(0)}]\,{E}^{(0)}_{++}\ , &
{\hat F}^{(1)}_{++} &= \frac12 [F^{(0)}_{++}-A_{++}^{(0)}]^2\ . \label{combTLpp}\\
{\hat A}^{(1)}_{+-}&=\frac12 [A^{(0)}_{+-}-C_{+-}^{(0)}]^2+\frac12 {B}^{(0)}_{+-} {F}^{(0)}_{+-} \ , &
{\hat B}^{(1)}_{+-}&=\frac12 [A^{(0)}_{+-}+E_{+-}^{(0)}-2 C_{+-}^{(0)}]\,{B}^{(0)}_{+-}\ ,\nonumber \\
{\hat C}^{(1)}_{+-}&=0\ ,&
{\hat D}^{(1)}_{+-}&=\frac12[D^{(0)}_{+-}-C_{+-}^{(0)}]^2\ ,\nonumber \\
{\hat E}^{(1)}_{+-}&=\frac12 [E^{(0)}_{+-}-C_{+-}^{(0)}]^2+\frac12 {B}^{(0)}_{+-} {F}^{(0)}_{+-} \ ,&
{\hat F}^{(1)}_{+-}&=\frac12 [A^{(0)}_{+-}+E_{+-}^{(0)}-2 C_{+-}^{(0)}]\,{F}^{(0)}_{+-}\ . \label{combTLpm}
\end{align}
The validity of these relations is rather general and can be applied to any
S-matrix with the same underlying structure. In particular, this allows us to
use them for the mixed flux case in section \ref{mixed}.

\subsubsection{The t-channel contribution and the dressing phases}\label{sec:tchannel}

As explained in section \ref{sec:gp} the t-channel cut requires a non-trivial
generalization of the procedure used for the $AdS_5\times S^5$ case
\cite{Bianchi:2013nra}. Furthermore, the t-channel cut for the scattering of
two masses depends on the tree-level S-matrices for the scattering of the same
and different masses. Therefore, in this section it only makes sense to work
with the parameters $\b$, $m$ and $m'$ for the three cases of interest, as
given in table \ref{assign}. Inputting the tree-level S-matrices
\eqref{treepp}, \eqref{treepm}, \eqref{tree3}, \eqref{tree4}, \eqref{tree5} and
\eqref{tree6} into eq~.\eqref{resultsummt} and splitting the result as in
eq.~\eqref{tchannelid} we find for all three scattering processes ($AdS_3
\times S^3 \times T^4$, $AdS_3 \times S^3 \times S^3 \times S^1$ same mass and
$AdS_3 \times S^3 \times S^3 \times S^1$ different mass) the one-loop phases
can be written in the following general form
\begin{align}
\phi^{(1)}_{++}(p,p')&=\frac{p\, p' (m'p+mp')^2}{8 \p \,m m' (e' p-e p') }\ ,\label{eq:phasepp}\\
\phi^{(1)}_{+-}(p,p')&=-\frac{p\, p' (m'p-mp')^2}{8 \p \, m m' (e' p-e p') }\ .\label{eq:phasepm}
\end{align}
The real part of the one-loop cut-constructible S-matrix that is not part of
the overall phase factors is given by
\begin{equation}\label{eq:counterterm}
\text{Re}(\hat{T}^{(1)})|_{\textup{unit.}}=\frac{1}{4\p} \, |1-2\,\b|\left(\frac{{p}^2}{m} +\frac{p'^2}{m'}\right)\, T^{(0)}\ .
\end{equation}
It is important to emphasise that even though we have written them in terms of
$\b$, $m$ and $m'$ the results \eqref{eq:phasepp}, \eqref{eq:phasepm} and
\eqref{eq:counterterm} are only valid for the assignments in table
\ref{assign}.

Two comments are in order here. First, eq.~\eqref{eq:counterterm} is
proportional to $|1-2\b|$. Therefore, this term vanishes for $AdS_3\times S^3
\times T^4$, but does not for $AdS_3\times S^3 \times S^3\times S^1$. However,
we should recall that this is only the contribution to
$\text{Re}(\hat{T}^{(1)})$ coming from unitarity and there are potentially
additional terms arising from external leg corrections \eqref{tchannelid}.
Indeed, one of the main differences between $AdS_3\times S^3 \times T^4$ and
$AdS_3\times S^3 \times S^3\times S^1$ is that the light-cone gauge-fixed
Lagrangian of the latter has cubic terms. Furthermore, the tree-level form
factor for one off-shell and two on-shell particles is non-zero and as a
consequence non-trivial external leg corrections are already present at one
loop in the unitarity construction, as described in section \ref{sec:elc}. As
we will see in the following section these precisely cancel
\eqref{eq:counterterm} and re-establish agreement with the exact
result.\footnote{Let us point out that a term like \eqref{eq:counterterm} in
the one-loop S-matrix would prevent the latter from satisfying the Yang-Baxter
equation, conflicting with the integrability of the theory.\vspace{2pt}}$^{,}$\footnote{It
is interesting to note that in the two loop near-flat-space computation of
\cite{Klose:2007rz} for the $AdS_5 \times S^5$ light-cone gauge S-matrix the
external leg corrections also cancelled unwanted terms arising from t-channel
graphs.\vspace{2pt}}

The second comment concerns eqs.~\eqref{lppS1}, \eqref{lpmS1},
\eqref{eq:phasepp} and \eqref{eq:phasepm}, which combined have a natural
interpretation as the one-loop contributions to the phases. It is interesting
to note that they are independent of $\b$, indicating that the phases for all
three scattering processes should be related. This agrees with the
semiclassical computation \cite{Abbott:2013kka}.\footnote{In
\cite{Abbott:2013ixa} the author states that the one-loop dressing phase of
$AdS_3\times S^3 \times S^3\times S^1$ is half that of $AdS_3\times S^3 \times
T^4$. This is consistent given that we are considering the factorized S-matrix
for $AdS_3\times S^3 \times T^4$.\vspace{2pt}} A natural question is whether this relation
extends to all orders in the coupling. Although this goes beyond the scope of
this work, to facilitate comparison with the literature \cite{Borsato:2013hoa}
we will rewrite the result in terms of the standard strong coupling variables
$x$ and $y$, which we have defined in \eqref{xexp} and \eqref{xofp}
\begin{align}
\vp^{(1)}_{++}(p,p')&=-\frac{m m'}{\pi}\frac{x^2}{x^2-1}\frac{y^2}{y^2-1}\bigg[\frac{(x+y)^2 (1-\frac{1}{xy})}{(x^2-1)(x-y)(y^2-1)} + \frac{2}{(x-y)^2} \log\left(\frac{x+1}{x-1}\frac{y-1}{y+1}\right)\bigg] \ , \label{1Lphasepp} \\
\vp^{(1)}_{+-}(p,p')&=-\frac{m m'}{\pi}\frac{x^2}{x^2-1}\frac{y^2}{y^2-1}\bigg[\frac{(xy+1)^2 (\frac{1}{x}-\frac{1}{y})}{(x^2-1)(xy-1)(y^2-1)} + \frac{2}{(xy-1)^2} \log\left(\frac{x+1}{x-1}\frac{y-1}{y+1}\right)\bigg] \label{1Lphasepm} \ .
\end{align}
Here $x$ corresponds to momentum $p$ with mass $m$ and $y$ to momentum $p'$
with mass $m'$. Finally, let us stress again that this expression is valid for
all three cases summarized in table \ref{assign}. In particular, for $m=m'=1$
this is consistent with \eqref{AFSstrong}, where the overall sign is
compensated by the fact that $e^{i\vartheta_{\s_M\s_N}(p,p')}\sim
S_{\s_M\s_N}^{11}(p,p')^{-1}$, see eqs.~\eqref{Spp} and \eqref{Spm}.

\subsubsection{External leg corrections for \txpf{$AdS_3\times S^3 \times S^3 \times S^1$}{AdS3 x S3 x S3 x S1}}\label{extlegs}

In this section we focus on the $AdS_3\times S^3 \times S^3 \times S^1$
background for which the unwanted term \eqref{eq:counterterm} is present. With
the aim of interpreting this missing term as a contribution cancelled by
external leg corrections let us review the results of \cite{Sundin:2012gc,
Sundin:2014sfa} for the one-loop two-point functions. The near-BMN expansion of
the light-cone gauge-fixed Lagrangian can be schematically written as
\begin{equation}
\mathcal{L}=\mathcal{L}_2+h^{-\frac12}\mathcal{L}_3+h^{-1}\mathcal{L}_4 + \ldots \ .
\end{equation}
The quadratic part is given by\,\footnote{Here we stress that, although the
theory is not Lorentz invariant beyond quadratic order, we are formally
rearranging the fermions into doublets for notational and computational
convenience.\vspace{2pt}}
\begin{align}
\mathcal{L}_2&= \bar{\chi}^a(i\slashed{\pa}-m_a) \chi^a+|\pa \varphi_a|^2-m_a^2 |\varphi_a|^2 \ ,
\end{align}
where our conventions are summarized in appendix \ref{notations} and we have
introduced the index $a=1,\ldots,4$ with the respective masses listed in table
\ref{masses}. The cubic Lagrangian \cite{Sundin:2012gc,Sundin:2014sfa} is given
by
\begin{align}\label{cublag}
\mathcal{L}_3= \sqrt{\frac{\a \bar\a}{2}}&\Big[
{(\chi^1)}{}^T \gamma^3 (i \slashed{\pa}-\a)\, \varphi_2\, \chi^3 
-i{(\chi^1)}{}^T \gamma^3 (i \slashed{\pa}-\bar\a)\, \varphi_3\, \chi^2 
-2{(\chi^2)}{}^T \gamma^1\pa_1 \varphi_1\, \chi^3 \nonumber \\
& \quad +{\bar{\chi}}^2\gamma^0(i \slashed{\pa}-\a)\, \varphi_2\, \chi^4  \vphantom{\Big]}
+i {\bar{\chi}}^3\gamma^0(i \slashed{\pa}-\bar\a)\, \varphi_3\, \chi^4 \nonumber \\
& \qquad - \left(\bar{\chi}^2 (1-\gamma^3) \chi^2-\bar{\chi}^3(1-\gamma^3) \chi^3
+ 2 \a|\varphi_2|^2-2\bar\a |\varphi_3|^2\right)\pa_0 \varphi_4+\text{h.c.} \Big]\ .
\end{align}

Let us start by focusing on the tree-level processes following from the cubic
Lagrangian. The only processes allowed by two-dimensional kinematics involve a
particle of mass 1 decaying into a particle of mass $\a$ and one of mass $\bar\a$
and its reverse.\footnote{Diagrams involving one massless leg are ruled out by
two-dimensional kinematics. In the cubic Lagrangian \eqref{cublag} the massless
modes always couple to massive modes of equal mass. It then follows that
the on-shell condition implies that the massless leg carries vanishing
momentum.\vspace{2pt}} The Feynman rules associated to the relevant vertices are
\begin{align}\nonumber
&\begin{tikzpicture}[scale=1,vertex/.style={circle,fill=black,thick,inner sep=2pt}]
\node (a1) at (-2,0) {$\varphi_1$};
\node (c) at (0,0) [vertex] {};
\node (a2) at (1.5,1) {$\chi^2$};
\node (a3) at (1.5,-1) {$\chi^3$};
\draw[dashed] (a1)--(c);
\begin{scope}[decoration={markings,mark = at position 0.5 with {\arrow[scale=1.4]{latex}}}]
\draw[postaction={decorate}] (a2)--(c);
\draw[postaction={decorate}] (a3)--(c);
\end{scope}
\end{tikzpicture}
&\quad&
\begin{tikzpicture}[scale=1,vertex/.style={circle,fill=black,thick,inner sep=2pt}]
\node (a1) at (-2,0) {$\chi^1$};
\node (c) at (0,0) [vertex] {};
\node (a2) at (1.5,1) {$\varphi_2$};
\node (a3) at (1.5,-1) {$\chi^3$};
\draw[dashed] (a2)--(c);
\begin{scope}[decoration={markings,mark = at position 0.5 with {\arrow[scale=1.4]{latex}}}]
\draw[postaction={decorate}] (a1)--(c);
\draw[postaction={decorate}] (a3)--(c);
\end{scope}
\end{tikzpicture}
&\quad&
\begin{tikzpicture}[scale=1,vertex/.style={circle,fill=black,thick,inner sep=2pt}]
\node (a1) at (-2,0) {$\chi^1$};
\node (c) at (0,0) [vertex] {};
\node (a2) at (1.5,1) {$\chi^2$};
\node (a3) at (1.5,-1) {$\varphi_3$};
\draw[dashed] (a3)--(c);
\begin{scope}[decoration={markings,mark = at position 0.5 with {\arrow[scale=1.4]{latex}}}]
\draw[postaction={decorate}] (a2)--(c);
\draw[postaction={decorate}] (a1)--(c);
\end{scope}
\end{tikzpicture}
\\
&\quad \sqrt{\frac{\a \bar\a}{2}}\, 2\,i\, \g^1 p_1 \ ,
&\quad& \quad -\sqrt{\frac{\a \bar\a}{2}}\, i\, \g^3 (\slashed{\rmp}_2+\a) \ ,
&\quad& \quad -\sqrt{\frac{\a \bar\a}{2}}\, \g^3 (\slashed{\rmp}_3+\bar\a)\ .\label{eq:decayprocess}
\end{align}
To obtain the amplitude one should contract the external legs with the fermion
polarizations and enforce the on-shell condition. The three diagrams share the
same on-shell kinematics, i.e. denoting the incoming momentum of the heavy
particle (with mass $m_1=1$) as $\rmp_1$, the outgoing momenta of the light
particles are given by $\rmp_2=\tfrac{m_2}{m_1} \rmp_1$ and $\rmp_3=\tfrac{m_3}{m_1} \rmp_1$, where
$m_3=m_1-m_2$.\footnote{This is true under the assumption of a relativistic
dispersion relation, which in this case holds just at tree level.\vspace{2pt}} Using the
property that $v(m_i \rmp_1)=\sqrt{m_i}v(\rmp_1)$ (see eq.~\eqref{polv}), it is
clear that both the second and the third diagrams vanish as
$(\slashed{\rmp}+1)v(\rmp) = 0$. Furthermore, the first diagram is also
identically zero as a consequence of the identity $v(\rmp)^T \g^1 v(\rmp)=0$.

One may ask how this is compatible with the result of \cite{Sundin:2012gc}
where the authors find a non-vanishing expression for the one-loop correction
to the propagators coming from the graph formed of two three-point vertices.
Focusing on the one-loop contribution to the self-energy of the heavy boson the
result of \cite{Sundin:2012gc} reads
\begin{equation}\label{selfenergy}
\Sigma^{(1)}_0(p) = i\braket{\varphi_1 \bar \varphi_1}^{(1)}=\frac{1}{\pi^2}(\a \log\a+\bar\a \log\bar\a)\, p^2\ . 
\end{equation}
This result is obtained setting $\rmp^2=1$ (i.e. putting the propagator
on-shell) and its dependence on $p$ is a consequence of the lack of Lorentz
invariance. In a unitarity computation with the setup described in section
\ref{sec:elc} the two tree-level form factors appearing in figure \ref{2ptfun}
would be vanishing in the strict on-shell limit and this contribution would not
be caught. However, as discussed in section \ref{sec:elc}, our treatment
ignored any kind of tadpole diagram contributing to the external leg
corrections. Moreover, as pointed out in \cite{Sundin:2012gc} the contribution
\eqref{selfenergy} can be understood as the one-loop term in the expansion of
$\hh(h)$. Combining this observation with the fact that, in all the examples we
have considered, ignoring tadpole diagrams gives the S-matrix up to corrections
in $\hh(h)$ we may argue that these are coming from tadpole diagrams whose
analysis would require the introduction of a regularization (see also
\cite{Sundin:2012gc}). This is therefore an additional indication that
unitarity techniques, neglecting tadpoles, are blind to shifts in the coupling.

Therefore, we will consider the following alternative question. Are there are
external leg corrections that are caught by unitarity and which are relevant
for the one-loop calculation? In the S-matrix computation we consider
scattering processes for which the external legs have masses $\a$ or $\bar\a$.
Therefore, the external leg corrections we compute come from diagrams similar
to the first graph in figure \ref{graphs} with masses $m_1=1$ and $m_2=\a$ or 
$m_2=\bar\a$.

We start by considering an external leg of mass $\a$. Using the vertices in
eq.~\eqref{eq:decayprocess} we find the following form factors
{\allowdisplaybreaks
\begin{align}
&\begin{tikzpicture}[scale=1,vertex/.style={circle,fill=black,thick,inner sep=2pt},baseline=-5pt]
\node (a1) at (-2,0) {$\varphi_2$};
\node (c) at (0,0) [vertex] {};
\node (a2) at (1.5,1) {$\chi^1$};
\node (a3) at (1.5,-1) {$\chi^3$};
\draw[dashed] (a1)--(c);
\begin{scope}[decoration={markings,mark = at position 0.5 with {\arrow[scale=1.4]{latex}}}]
\draw[postaction={decorate}] (a2)--(c);
\draw[postaction={decorate}] (a3)--(c);
\end{scope}
\begin{scope}[decoration={markings,mark = at position 1 with {\arrow[scale=1.4]{to}}}]
\draw[postaction={decorate}] (-1.3,-0.25)--(-1,-0.25) node [anchor=north] {$\rmp$}--(-0.7,-0.25);
\draw[postaction={decorate}] (0.2,1/3+0.2)--(0.45,1/2+0.2) node [anchor=south] {$\rml_1$}--(0.7,2/3+0.2);
\draw[postaction={decorate}] (0.7,-2/3-0.2)--(0.45,-1/2-0.2) node[anchor=north] {$\rml_2$}--(0.2,-1/3-0.2);
\end{scope}
\end{tikzpicture} \quad = \quad i\,\sqrt{\frac{\a \bar\a}{2}}\, v(\rml_1)^{T} \gamma^3 (\slashed{\rmp}-\a) u(\rml_2)\ ,\label{diagscalar}
\\ &
\begin{tikzpicture}[scale=1,vertex/.style={circle,fill=black,thick,inner sep=2pt},baseline=-5pt]
\node (a1) at (-2,0) {$\chi^2$};
\node (c) at (0,0) [vertex] {};
\node (a2) at (1.5,1) {$\varphi_1$};
\node (a3) at (1.5,-1) {$\chi^3$};
\draw[dashed] (a2)--(c);
\begin{scope}[decoration={markings,mark = at position 0.5 with {\arrow[scale=1.4]{latex}}}]
\draw[postaction={decorate}] (a1)--(c);
\draw[postaction={decorate}] (a3)--(c);
\end{scope}
\begin{scope}[decoration={markings,mark = at position 1 with {\arrow[scale=1.4]{to}}}]
\draw[postaction={decorate}] (-1.3,-0.25)--(-1,-0.25) node [anchor=north] {$\rmp$}--(-0.7,-0.25);
\draw[postaction={decorate}] (0.2,1/3+0.2)--(0.45,1/2+0.2) node [anchor=south] {$\rml_1$}--(0.7,2/3+0.2);
\draw[postaction={decorate}] (0.7,-2/3-0.2)--(0.45,-1/2-0.2) node[anchor=north] {$\rml_2$}--(0.2,-1/3-0.2);
\end{scope}
\end{tikzpicture} \quad = \quad i\, \sqrt{\frac{\a \bar\a}{2}}\, 2\, u(\rml_2)^{T}\g^1 l_1 u(\rmp)\ ,\label{diagfermion1}
\\ &
\begin{tikzpicture}[scale=1,vertex/.style={circle,fill=black,thick,inner sep=2pt},baseline=-5pt]
\node (a1) at (-2,0) {$\chi^2$};
\node (c) at (0,0) [vertex] {};
\node (a2) at (1.5,1) {$\chi^1$};
\node (a3) at (1.5,-1) {$\varphi_3$};
\draw[dashed] (a3)--(c);
\begin{scope}[decoration={markings,mark = at position 0.5 with {\arrow[scale=1.4]{latex}}}]
\draw[postaction={decorate}] (a2)--(c);
\draw[postaction={decorate}] (a1)--(c);
\end{scope}
\begin{scope}[decoration={markings,mark = at position 1 with {\arrow[scale=1.4]{to}}}]
\draw[postaction={decorate}] (-1.3,-0.25)--(-1,-0.25) node [anchor=north] {$\rmp$}--(-0.7,-0.25);
\draw[postaction={decorate}] (0.2,1/3+0.2)--(0.45,1/2+0.2) node [anchor=south] {$\rml_1$}--(0.7,2/3+0.2);
\draw[postaction={decorate}] (0.7,-2/3-0.2)--(0.45,-1/2-0.2) node[anchor=north] {$\rml_2$}--(0.2,-1/3-0.2);
\end{scope}
\end{tikzpicture} \quad = \quad \sqrt{\frac{\a \bar\a}{2}}\, v(\rml_1)^{T}\g^3(\slashed{\rml}_2-\bar\a)u(\rmp)\ .\label{diagfermion2}
\end{align}}

In order to apply the construction outlined in section \ref{sec:elc} we need to
compute eq.~\eqref{ffsquare}. In particular, we are interested in expanding the
form factor squared around the on-shell condition. Since we already know that
the tree-level form factor vanishes on-shell, to get the first order in the
expansion there is no need to also expand the integral, i.e. it can be
evaluated strictly on-shell
\begin{align}
I(\a^2,1,\bar\a)=-\frac{i}{4\p\bar\a}\ .
\end{align}
Squaring the form factor \eqref{diagscalar} and expanding around the on-shell
condition we find\,\footnote{A minus sign is included to take account of the
fermion loop.\vspace{2pt}}
\begin{equation}\label{eq:scalarext}
-i\Sigma^{(1)}_{1,\vf_2}(\rmp)=\frac{i}{4\p\a}\, p^2\ .
\end{equation}
Comparing to \eqref{eq:counterterm} this result is promising. However,
\eqref{eq:scalarext} holds only when the external leg is a boson. A non-trivial
check of our procedure is that when the external leg is a fermion the
correction, which comes from two terms associated to the diagrams
\eqref{diagfermion1} and \eqref{diagfermion2}, is exactly the same as for the
boson, i.e.
\begin{equation}\label{eq:fermionext}
-i\Sigma^{(1)}_{1,\chi^2}(\rmp)=\frac{i}{4\p\a}\, p^2\ .
\end{equation}
One might have expected this from worldsheet supersymmetry as discussed in
\cite{Sundin:2014sfa}. Here we have computed the external leg corrections for a
particle of mass $\a$. From the symmetry of the Lagrangian, it is clear
that the result for a particle of mass $\bar\a$ is just given by the
replacement $\a\to\bar\a$.

Once the external leg contributions are computed we can apply
eq.~\eqref{important} to find their contribution to the one-loop S-matrix. To
be general, let us consider the scattering of a particle of mass $m$ with a
particle of mass $m'$. Our result then reads
\begin{equation}
T^{(1)}_{ext}=-\frac{1}{4\pi}\bigg(\frac{p^2}{m}+\frac{p'^2}{m'}\bigg) T^{(0)}\ .
\end{equation}
This contribution exactly cancels \eqref{eq:counterterm} for $\beta=0$ and
$\beta=1$. These are precisely the values associated to the single and mixed
mass scattering processes for $AdS_3\times S^3 \times S^3 \times S^1$, and
hence we have established agreement between the unitarity calculation and the
exact result up to shifts in the coupling.

\section{Massive sector for \txpf{$AdS_3\times S^3 \times T^4$}{AdS3 x S3 x T4} supported by mixed flux}\label{mixed}

In this section we apply the methods of section \ref{sec:gp} to the massive
sector of the $AdS_3 \times S^3 \times T^4$ light-cone gauge-fixed string
theory supported by a mix of RR and NSNS fluxes. We parameterize the relative
strength of the two fluxes with a parameter $q$, with $q=0$ corresponding to
pure RR and $q=1$ to pure NSNS flux. The symmetry algebras underlying this
theory and its S-matrix are the same for all $q$
\cite{Pesando:1998wm,Cagnazzo:2012se,Hoare:2013pma,Hoare:2013ida} -- the
parameter $q$ enters through the form of the representation of the algebra on
the states.

\subsection{Exact S-matrix and tree-level result}

The quadratic light-cone gauge-fixed action for the $AdS_3 \times S^3 \times
T^4$ background supported by mixed flux again describes $4+4$ massive and $4+4$
massless fields. As usual we restrict ourselves to considering the scattering
of two massive excitations to two massive excitations. Following the RR case
described in section \ref{sec:t4} we group the particle content of the massive
sector into $2+2$ complex degrees of freedom (to recall, $\Phi_{\vf\vf}$,
$\Phi_{\psi\psi}$, $\Phi_{\vf\psi}$, $\Phi_{\psi\vf}$, and their complex
conjugates $\Phi_{\bar \vf\bar\vf}$, $\Phi_{\bar \psi \bar \psi}$, $\Phi_{\bar
\vf\bar \psi}$, $\Phi_{\bar \psi\bar \vf}$). The presence of the NSNS flux
then breaks the charge conjugation invariance, such that the near-BMN
dispersion relations for these complex degrees of freedom are given by
\begin{equation}\label{eq:dr}
e_\pm = \sqrt{(1 -q^2) + (p \pm q)^2} \ .
\end{equation}
where $+$ corresponds to $\Phi_{\vf\vf}$, $\Phi_{\psi\psi}$, $\Phi_{\vf\psi}$,
$\Phi_{\psi\vf}$ and $-$ to their complex conjugates.

As for $q=0$ the S-matrix factorizes as in \eqref{factorize} and the general
structure of the factorized S-matrix takes the form given in \eqref{structure}
with $\s_\vf=\s_\psi=+$ and $\s_{\bar{\vf}}=\s_{\bar{\psi}}=-$. Furthermore,
the construction outlined in section \ref{sec:gp} still gives the same one-loop
result whether we consider the factorized or full S-matrix. Therefore, for
simplicity we will again work with the former. Due to the lack of charge
conjugation symmetry all four phases are now different.  However, charge
conjugation along with formally sending $q \to -q$ is a symmetry and hence
$\phi_{++}=\phi_{--}|_{q\to -q}$ and $\phi_{+-}=\phi_{-+}|_{q\to -q}$.
Similarly, for the functions $\cl_{\s_M\s_N}$ we have $\cl_{++}=\cl_{--}|_{q\to
-q}$ and $\cl_{+-}=\cl_{-+}|_{q\to -q}$. Therefore, in the following we will
again focus on the $++$ and $+-$ sectors.
The dependence on the gauge-fixing parameter $a$ is also modified in the
following natural way
\begin{equation}\label{qgauge}
\exp\big[\frac{i}{2}(a-\tfrac12)(\e_{\s_N}'p-\e_{\s_M} p')\big] \ ,
\end{equation}
where the all-order energies $\e_\pm$ are defined in appendix \ref{notations2}.
As discussed in section \ref{resstruc} we choose the overall phase factors by
setting particular components of $\hat{S}_{MN}^{PQ}$ to one
\begin{align}
\hat{S}_{\vf\vf}^{\vf\vf}(p,p')&=1\ , \qquad \hat{S}_{\vf\bar{\psi}}^{\vf\bar{\psi}}(p,p')=1\ . \label{qchoice}
\end{align}

The parametrizing functions of the exact S-matrix are defined as
\begin{align}
{S}_{\vf\vf}^{\vf\vf}(p,p')&=A_{++}(p,p') & {S}_{\vf\bar{\vf}}^{\vf\bar{\vf}}(p,p')&=A_{+-}(p,p')\nonumber\\
{S}_{\vf\psi}^{\vf\psi}(p,p')&=B_{++}(p,p') & {S}_{\vf\bar{\vf}}^{\psi\bar{\psi}}(p,p')&=B_{+-}(p,p')\nonumber\\
{S}_{\vf\psi}^{\psi\vf}(p,p')&=C_{++}(p,p') & {S}_{\vf\bar{\psi}}^{\vf\bar{\psi}}(p,p')&=C_{+-}(p,p')\nonumber\\
{S}_{\psi\vf}^{\psi\vf}(p,p')&=D_{++}(p,p') & {S}_{\psi\bar{\vf}}^{\psi\bar{\vf}}(p,p')&=D_{+-}(p,p')\nonumber\\
{S}_{\psi\vf}^{\vf\psi}(p,p')&=E_{++}(p,p') & {S}_{\psi\bar{\psi}}^{\psi\bar{\psi}}(p,p')&=E_{+-}(p,p')\nonumber\\
{S}_{\psi\psi}^{\psi\psi}(p,p')&=F_{++}(p,p') & {S}_{\psi\bar{\psi}}^{\vf\bar{\vf}}(p,p')&=F_{+-}(p,p')\label{qcorr}
\end{align}
with the functions in string frame given by \cite{Hoare:2013ida}
\begin{align}
A_{++}(p,p') &= S_{++}(p,p') \ , &
B_{++}(p,p') &= S_{++}(p,p') \frac{x'^+_+ - x^+_+}{x'^+_+ - x^-_+} \frac{1}{\n_+}\ , \nonumber \\
C_{++}(p,p') &= S_{++}(p,p') \frac{x'^+_+ - x'^-_+}{x'^+_+ - x^-_+} \frac{\eta_+}{\eta'_+} \sqrt{\frac{\n'_+}{\n_+}}\ , &
D_{++}(p,p') &= S_{++}(p,p') \frac{x'^-_+ - x^-_+}{x'^+_+ - x^-_+} \n'_+\ , \nonumber \\
E_{++}(p,p') &= S_{++}(p,p') \frac{x^+_+ - x^-_+}{x'^+_+ - x^-_+} \frac{\eta'_+}{\eta_+} \sqrt{\frac{\n'_+}{\n_+}}\ , &
F_{++}(p,p') &= S_{++}(p,p') \frac{x'^-_+ - x^+_+}{x'^+_+ - x^-_+}\frac{\n'_+}{\n_+}\ , \label{qfunpp}
\end{align}
and
\begin{align}
A_{+-}(p,p') &= S_{+-}(p,p') \frac{1-\frac{1}{x^+_+\, x'^-_-}}{1-\frac{1}{x^-_+\, x'^-_-}} \n_+\ , &
B_{+-}(p,p') &= -S_{+-}(p,p') \frac{i\, \eta_+\eta'_-}{{x}^-_+ x'^-_-}\frac{1}{1-\frac{1}{x^-_+\, x'^-_-}} (\n_+\n'_-)^{-\frac12}\ , \nonumber \\
C_{+-}(p,p') &= S_{+-}(p,p') \ , &
D_{+-}(p,p') &= S_{+-}(p,p') \frac{1-\frac{1}{x^+_+\, x'^+_-}}{1-\frac{1}{x^-_+\, x'^-_-}}\n_+\n'_-\ , \nonumber \\
E_{+-}(p,p') &= S_{+-}(p,p') \frac{1-\frac{1}{x^-_+\, x'^+_-}}{1-\frac{1}{x^-_+\, x'^-_-}} \n'_-\ , &
F_{+-}(p,p') &= -S_{+-}(p,p') \frac{i\, \eta_+\eta'_-}{x^+_+ x'^+_-}\frac{1}{1-\frac{1}{x^-_+\, x'^-_-}}(\n_+\n'_-)^\frac32\ . \label{qfunpm}
\end{align}
The definitions of the variables entering these expressions are given in
appendix \ref{notations2}. The functions $S_{++}(p,p')$ and $S_{+-}(p,p')$ are
two of the overall phase factors. These phase factors are not fixed by
symmetry -- they are constrained by crossing symmetry, however, they are
currently unknown.

The input needed for the unitarity construction of section \ref{sec:gp} is the
tree-level S-matrix. Various tree-level components were computed directly in
\cite{Hoare:2013pma}. These are in agreement with the near-BMN expansion of the
exact result \eqref{qfunpp}, \eqref{qfunpm}, for which we recall that the
integrable coupling used in the definition of $x^\pm_+$ and $x^\pm_-$ has the
expansion $\hh (h)=h +\mathcal{O}(h^0)$ and to take the near-BMN limit the
spatial momenta should be rescaled $p\to \frac{p}{h}$. The remaining
components of the tree-level S-matrix can then be fixed from the expansion of
the exact result. As in the RR case, here we shall present the result in the
gauge $a=\frac12$ -- the dependence on $a$ goes through the unitarity
procedure without any particular subtlety, i.e. it exponentiates as in
eq.~\eqref{qgauge}. The tree-level S-matrix reads \begin{align}
A^{(0)}_{++}(p,p') &= - F^{(0)}_{++}(p,p') = \frac{(p+p')(e'_+p+e_+p')}{4\,(p-p')}\ , \qquad \nonumber \\
C^{(0)}_{++}(p,p') &= E^{(0)}_{++}(p,p') = \frac{p\, p'}{2(p-p')}\big(\sqrt{(e_++p+q)(e'_++p'+q)}+\sqrt{(e_+-p-q)(e'_+-p'-q)}\big)\ , & \nonumber \\
B^{(0)}_{++}(p,p') &= - D^{(0)}_{++}(p,p') = -\frac{e'_+ p-e_+ p'}4\ ,\label{qtrepp}
\end{align}
\begin{align}
A^{(0)}_{+-}(p,p') &= - E^{(0)}_{+-}(p,p') = \frac{(p-p')(e'_-p+e_+p')}{4\,(p+p')}\ , \qquad \nonumber \\
B^{(0)}_{+-}(p,p') &= F^{(0)}_{+-}(p,p') = \frac{p\, p'}{2(p+p')}\big(\sqrt{(e_+-p-q)(e'_--p'+q)}-\sqrt{(e_++p+q)(e'_-+p'-q)}\big)\ , & \nonumber \\
C^{(0)}_{+-}(p,p') &= - D^{(0)}_{+-}(p,p') = -\frac{e'_- p-e_+ p'}4\ ,
\label{qtrepm}
\end{align}
This form of writing the tree-level S-matrix elements is the simplest for the
purposes of introducing the parameter $q$. Agreement with \eqref{treepp} and
\eqref{treepm} for $q=0$ can be checked using the dispersion relation.

\subsection{Result from unitarity techniques}\label{sec:phq}

In this section we compute the one-loop S-matrix from unitarity methods for the
light-cone gauge-fixed string theory in the $AdS_3 \times S^3 \times T^4$
background supported by a mix of RR and NSNS fluxes. Again, we will split the
result according to eqs.~\eqref{logsid}, \eqref{schannelid} and
\eqref{tchannelid}, where we recall that we have chosen
$S^{\vf\vf}_{\vf\vf}=A_{++}(p,p')$ and $S^{\vf \bar\psi}_{\vf\bar\psi}=C_{+-}(p,p')$
as the overall phase factors.

There is a subtlety regarding the unitarity computation in that the near-BMN
dispersion relations \eqref{eq:dr} are not the standard relativistic ones that
we assumed for the derivation in section \ref{sec:gp}. To bypass this problem,
we will first shift the momenta as
\begin{equation} \begin{split}\label{shift}
& p \to p - q \text{ for particles and } p \to p + q \text{ for antiparticles} \ ,
\end{split} \end{equation}
so as to put the near-BMN dispersion relations into the standard form.
At the level of the light-cone gauge-fixed Lagrangian this just amounts to a
$\s$-dependent rotation of the complex fields, where $\s$ is the spatial
coordinate on the worldsheet \cite{Hoare:2013pma}. We can then
straightforwardly use the construction of section \ref{sec:gp} for two
particles of mass $\sqrt{1-q^2}$. To construct the one-loop result, we should
then conclude by undoing the shift \eqref{shift}. An analogous approach was used
in \cite{Engelund:2013fja} to compute to logarithmic terms.

Following this procedure it is apparent that the logarithms appearing in the
one-loop integrals, when written in terms of energy and momentum, are different
for each of the four sectors
\begin{equation}
\theta_{\pm\pm} = \as\big(\frac{e'_\pm(p\pm q)-e_\pm(p'\pm q)}{1-q^2}\big)\ , \qquad
\theta_{\pm\mp} = \as\big(\frac{e'_\mp(p\pm q)-e_\pm(p'\mp q)}{1-q^2}\big)\ .
\end{equation}
The functions $\ell_{\s_M\s_N}$ are then defined as the coefficients of
$\theta_{\s_M\s_N}$ in the one-loop phase, see eq.~\eqref{phasestruc}.

\

We start by briefly reviewing the work of \cite{Engelund:2013fja}. Given that
the structure of the S-matrix is not altered by the presence of NSNS flux it
follows from the unitarity computation that the coefficients of the logarithms
written in terms of the tree-level functions, \eqref{qtrepp} and
\eqref{qtrepm}, are still given by \eqref{logs1} and \eqref{logs2}
\begin{align}
\cl_{++}(p,p')&=-\frac{1}{2\pi}{C}^{(0)}_{++}(p,p'){E}^{(0)}_{++}(p,p') =- \frac{p^2p'^2\big(e_+ e_+' + (p+q)(p'+q) +(1-q^2)\big)}{4\pi(p-p')^2} \ ,\label{qlogs1}\\
\cl_{+-}(p,p')&=-\frac{1}{2\pi}{B}^{(0)}_{+-}(p,p'){F}^{(0)}_{+-}(p,p') =- \frac{p^2p'^2\big(e_+ e_-' +(p+q)(p'-q)-(1-q^2)\big)}{4\pi(p+p')^2}\ .\label{qlogs2}
\end{align}
Using the dispersion relation, one can check that these expressions agree with
eqs.~\eqref{lppS1} and \eqref{lpmS1} for $q=0$ and $m=m'=1$.

Furthermore, the rational s-channel terms (with the overall phase factors set
to one) are again given in terms of the tree-level functions as in
eqs.~\eqref{combTLpp} and \eqref{combTLpm}. Plugging in the corresponding
expressions, \eqref{qtrepp} and \eqref{qtrepm}, one can check agreement with
the near-BMN expansion of the exact result \eqref{qfunpp} and \eqref{qfunpm}.

Finally, as for the $AdS_3 \times S^3 \times T^4$ background supported by pure
RR flux, the rational contributions from the t-channel go completely into the
phases. That is $\text{Re}(\hat{T}^{(1)})|_{\textup{unit.}} = 0$. Furthermore,
also as for the case of pure RR flux, the light-cone gauge-fixed Lagrangian
contains no cubic terms. Therefore, there are correspondingly no external leg
corrections at one loop in the unitarity computation. It follows from
computing the t-channel cuts that
\begin{align}
\phi^{(1)}_{++}(p,p')&=\frac{p\, p' (p+p')(e_+'p+e_+p')}{8 \p (p-p')}\ ,\label{eq:qphasepp}\\
\phi^{(1)}_{+-}(p,p')&=-\frac{p\, p' (p-p')(e_-'p+e_+p')}{8\p(p+p') }\ .\label{eq:qphasepm}
\end{align}
Using the dispersion relation, one can check that these expressions agree with
eqs.~\eqref{eq:phasepp} and \eqref{eq:phasepm} for $q=0$ and $m=m'=1$.

\

We conclude this section by giving the generalization of the one-loop dressing
phases \eqref{1Lphasepp} and \eqref{1Lphasepm} in the presence of NSNS flux. As
discussed in appendix \ref{notations2} the standard strong coupling variables
$x$ and $y$ are modified for $q\neq 0$. In particular, we now have a separate
variable for the particle $x_+$, $y_+$ and the antiparticle $x_-$, $y_-$. These
are defined in \eqref{xqexp} and \eqref{xqofp}. Our conjecture for the one-loop
dressing phases is then given by ($x_\pm$ corresponds to $p$ and $y_\pm$ to
$p'$)
\begin{align}
\vp^{(1)}_{++}(p,p')&=-\frac{1}{\pi}\frac{x_+^2}{\qh(x_+^2-1)-2q x_+}\frac{y_+^2}{\qh(y_+^2-1)-2qy_+}\nonumber
\\ & \qquad\quad\bigg[\frac{(x_++y_+)\big(\qh(x_++y_+)(1-\frac{1}{x_+y_+})-4q\big)}{(\qh(x_+^2-1)-2qx_+)(x_+-y_+)(\qh(y_+^2-1)-2qy_+)} \nonumber
\\ & \qquad \qquad \qquad + \frac{2}{(x_+-y_+)^2} \log\left(\frac{\qp\, x_++\qm}{\qm\, x_+-\qp}\frac{\qm\, y_+-\qp}{\qp \,y_++\qm}\right)\bigg] \ , \label{1Lphaseppq} \\
\vp^{(1)}_{+-}(p,p')&=-\frac{1}{\pi}\frac{x_+^2}{\qh(x_+^2-1)-2q x_+}\frac{y_-^2}{\qh(y_-^2-1)+2qy_-}\nonumber
\\ & \qquad \quad \bigg[\frac{(x_+y_-+1)\big(\qh(x_+y_-+1)(\frac{1}{x_+}-\frac{1}{y_-}) + 4q\big)}{(\qh(x_+^2-1)-2qx_+)(x_+-y_-)(\qh(y_-^2-1)+2q y_-))} \nonumber
\\ & \qquad \qquad \qquad + \frac{2}{(x_+y_--1)^2} \log\left(\frac{\qp\,x_++\qm}{\qm\,x_+-\qp}\frac{\qp\,y_--\qm}{\qm\,y_-+\qp}\right)\bigg] \label{1Lphasepmq} \ .
\end{align}

\section{Comments}

In this paper we have applied one-loop unitarity methods to the massive sectors
of three $AdS_3 \times S^3 \times M^4$ light-cone gauge-fixed string theories.
To do so we extended the construction of \cite{Bianchi:2013nra} to account for
both particles of different mass in the asymptotic spectrum and external leg
corrections. Applying the results to the $AdS_3 \times S^3 \times T^4$ and
$AdS_3 \times S^3 \times S^3 \times S^1$ backgrounds supported by RR flux
correctly reproduces the expansion of the exact S-matrix
\cite{Borsato:2013qpa,Borsato:2013hoa} and matches semiclassical computations
\cite{Abbott:2012dd,Beccaria:2012kb,Abbott:2013ixa} (see footnote 
\ref{clarification}). This agreement crucially included the rational terms,
supporting the conjecture of \cite{Bianchi:2013nra} that the S-matrices of
integrable field theories are cut-constructible (up to a possible shift in the
coupling).

The final theory we investigated was $AdS_3 \times S^3 \times T^4$ supported by
a mix of RR and NSNS fluxes. For this theory the one-loop unitarity computation
reproduced the exact result up to phases, which in this case are not known
beyond tree level. Under the assumption that the unitarity computation still
reproduces all the rational terms, we gave a conjecture for the one-loop dressing phases.

All three of these theories contain massless modes in their
spectrum. Recently an exact form for the S-matrix describing the scattering of
massless modes in the $AdS_3 \times S^3 \times T^4$ background was conjectured
in \cite{Borsato:2014exa}, and various perturbative computations have been
reported in \cite{Sundin:2013ypa}. It would therefore be of clear interest to
extend the unitarity methods to include these excitations.

Massless modes are of interest also for scattering on top of the GKP string
\cite{Gubser:1998bc,Basso:2013pxa}, for which excitations on the sphere
governed by the O(6) non-linear sigma model \cite{Alday:2007mf}. This S-matrix
recently turned out to be useful for studying $\mathcal{N}=4$ scattering
amplitudes in the collinear limit \cite{Basso:2013vsa,Basso:2014jfa}.

There are a number of further possible applications of the methods described in
this work (along with those in \cite{Bianchi:2013nra,Engelund:2013fja}). These
include the study of other integrable string backgrounds
\cite{Zarembo:2010sg,Wulff:2014kja} and more general off-shell objects,
including form factors \cite{Klose:2012ju,Klose:2013aza} and correlation
functions. Finally, one of the most important directions for future study is
the derivation of the rational terms from unitarity methods beyond one loop.
This would require a deeper understanding of both how to treat tadpole diagrams
and shifts in the coupling or lack thereof.

\textbf{Note added}: While this paper was in the final stages of preparation
the article \cite{Babichenko:2014yaa} was announced on arXiv. Among other
things, the authors reported on the semiclassical computation of the one-loop
phases for the $AdS_3 \times  S^3 \times T^4$ mixed flux background. The
conjecture for these phases given in this paper agrees with the result of
\cite{Babichenko:2014yaa}.

\section*{Acknowledgments}
We would like to thank V. Forini for fruitful discussions and enjoyable
collaborations on related work. We are also grateful to B. Basso, R. Borsato,
O. Engelund, S. Komatsu, M. Meineri, J. Plefka, R.  Roiban, A. Tseytlin and P.
Vieira for useful discussions and to R. Roiban and A. Tseytlin for valuable
comments on the draft. This work is funded by DFG via the Emmy Noether Program
``Gauge Fields from Strings.''. LB would like to thank Perimeter Institute for
Theoretical Physics for the kind hospitality during the preparation of this
work.

\setcounter{equation}{0}

\appendix

\section{Notation and conventions}\label{notations}

\subsection{\txpf{$AdS_3\times S^3\times M^4$}{AdS3 x S3 x M4} supported by RR flux}\label{notations1}

In section \ref{sec:exT4} the exact S-matrices are written as functions of the
Zhukovsky variables $x^{\pm}$ and $y^{\pm}$. These are defined in terms of the
energy and momentum as follows
\begin{align}
\frac{x^+}{x^-}&=e^{i p}\ , & x^+ - \frac{1}{x^+}-x^- + \frac{1}{x^-}&=\frac{2\,i\,\e}{\hh} \ , & x^+ + \frac{1}{x^+}-x^- - \frac{1}{x^-}&=\frac{2\,i\,m}{\hh} \ , \\
\frac{y^+}{y^-}&=e^{i p}\ , & y^+ - \frac{1}{y^+}-y^- + \frac{1}{y^-}&=\frac{2\,i\, \e}{\hh}\ ,& y^+ + \frac{1}{y^+}-y^- - \frac{1}{y^-}&=\frac{2\,i\,m'}{\hh}\ ,
\end{align}
where $\hh$ is the integrable coupling that is (potentially non-trivially)
related to the string tension $h$. The third equation of each line is a
constraint that is interpreted as the dispersion relation. In particular, $m$
and $m'$ are the masses of the respective particles. The variables $x'^{\pm}$
and $y'^{\pm}$ are simply given by sending $p\to p'$ and $\e \to \e'$.
Solving for $x^{\pm}$ and $y^{\pm}$ in terms of $p$ we find
\begin{align}
x^{\pm}&=\frac{e^{\pm i \frac{p}{2}}(m+\e)}{2\, \hh\, \sin\frac{p}{2}}\ , & \e&=\sqrt{m^2+4\,\hh^2\sin^2\frac{p}{2}}\ , \label{xpm}\\
y^{\pm}&=\frac{e^{\pm i \frac{p}{2}}(m'+\e)}{2\, \hh\, \sin\frac{p}{2}}\ , & \e&=\sqrt{m'^2+4\,\hh^2\sin^2\frac{p}{2}}\ . \label{ypm}
\end{align}
When expanding in near-BMN regime, the spatial momenta should first be rescaled
as $p\to \frac{p}{h}$ where $h$ is the string tension. The integrable coupling
$\hh$, in principle, is related to $h$ in a non-trivial way, however, its
strong coupling expansion starts with $\hh(h)= h + \mathcal{O}(h^0)$.
Therefore, at leading order in the near-BMN expansion the dispersion relation
is given by its relativistic counterpart, $e$. The two additional functions
that we use to write the expressions for the exact S-matrices are
\begin{equation}\label{etanu}
\eta=\sqrt{i(x^- - x^+)} \ , \qquad \n=\sqrt{\frac{x^+}{x^-}}\ ,
\end{equation}
and similarly for $y^\pm$ when referring to a particle of mass $m'$.

In section \ref{sec:tchannel} we are interested in expanding the functions
$x^\pm$ and $y^\pm$ at strong coupling. To do so it is convenient to introduce
a new variable $x$ such that
\begin{equation}\label{xexp}
x^\pm=x\pm\frac{im}{\hh} \frac{x^2}{x^2-1}+\mathcal{O}(\hh^{-3})\ .
\end{equation}
Expressing $x$ in terms of $p$ in the near-BMN expansion (i.e. first rescaling
$p$) one finds
\begin{equation}\label{xofp}
x(p)=\frac{m+\sqrt{m^2+p^2}}{p}+\mathcal{O}(h^{-2})\ .
\end{equation}
Using the new variable one can easily expand the dressing phase at strong
coupling as shown in appendix \ref{phases}.

In the discussion of the $AdS_3 \times S^3 \times S^3 \times S^1$ background we
need to use the cubic terms in the expansion of the light-cone gauge-fixed
Lagrangian. We use a worldsheet metric with signature $(+,-)$. Light-cone
coordinates are defined for a generic two-dimensional vector $v^\mu$ as
$v^{\pm}=\frac12(v^0\pm v^1)$ and for a covector $v_\mu$ as $v_{\pm}=v_0\pm
v_1$. The non-vanishing elements of the metric in light-cone coordinates are
$\eta_{+-}=\eta_{-+}=2$. Correspondingly $\eta^{+-}=\eta^{-+}=\frac12$. The
Levi-Civita tensor is defined as $\e^{01}=1=-\e_{01}$.

As usual, gamma matrices are defined by the anti-commutation relation
\begin{equation}
\{\c^\mu,\c^\nu\}=2\,\eta^{\mu\nu}\ .
\end{equation}
An explicit representation is given by
\begin{equation}
\c^0=\bigg(\begin{array}{cc}
0&1\\
1&0
\end{array}\bigg), \qquad
\c^1=\bigg(\begin{array}{cc}
0&1\\
-1&0
\end{array}\bigg), \qquad
\c^3=-\c^0\c^1=\bigg(\begin{array}{cc}
1&0\\
0&-1
\end{array}\bigg)\ .
\end{equation}
A generic spinor is represented as
\begin{equation}
\chi=\bigg(\begin{array}{c}
\chi_+\\
\chi_-
\end{array}\bigg) \ ,
\end{equation}
where $\chi_{\pm}$ are the chiral projections of $\chi$ by the projectors
$P_{\pm}=\frac12(1\pm\g^3)$. The conjugation is defined in the usual way
$\bar{\chi}=\chi^\dagger\g^0$ and to make contact with \cite{Sundin:2012gc,
Sundin:2014sfa} we define $\bar\chi_\pm\equiv\chi_\pm^\dagger$. The
polarization vectors can be chosen to be purely real and given by
\begin{align}
&\label{polu}\begin{tikzpicture}[baseline=-3pt]
\begin{scope}[decoration={markings,mark = at position 0.5 with {\arrow[scale=1.4]{latex}}}]
\draw[postaction={decorate}] (-1,0)--(0,0);
\end{scope}
\draw [pattern=crosshatch dots] (0.5,0) circle (0.5);
\end{tikzpicture}
\qquad u(\rmp) = \bigg(\begin{array}{c}\sqrt{p_+}\\\sqrt{p_-}\end{array}\bigg) \ ,
\\\nonumber \\
&\label{polv}\begin{tikzpicture}[baseline=-3pt]
\draw [pattern=crosshatch dots] (0.5,0) circle (0.5);
\begin{scope}[decoration={markings,mark = at position 0.5 with {\arrow[scale=1.4]{latex}}}]
\draw[postaction={decorate}] (2,0)--(1,0);
\end{scope}
\end{tikzpicture}
\qquad v(\rmp) = \bigg(\begin{array}{c}\sqrt{p_+}\\-\sqrt{p_-}\end{array}\bigg)\ .
\end{align}

\subsection{\txpf{$AdS_3\times S^3\times T^4$}{AdS3 x S3 x T4} supported by mixed flux}\label{notations2}

In the mixed flux case discussed in section \ref{mixed} the S-matrix is again
written in terms of Zhukovsky-type variables. However, the dispersion relation
is modified and is different for particles ($x^\pm_+$) and antiparticles
($x^\pm_-$). The Zhukovsky variables are defined in terms of the energy and
momentum as follows
\begin{align}
\frac{x^+_\spm}{x^-_\spm}&=e^{i p}\ , & x^+_\spm - \frac{1}{x^+_\spm}-x^-_\spm + \frac{1}{x^-_\spm}&=\frac{2\,i\,\e_\pm}{\hh\sqrt{1-q^2}}\ .
\end{align}
However, the dispersion relation \cite{Hoare:2013lja} is now given by
\begin{equation}
\sqrt{1-q^2}\big( x_\spm^+ + \frac{1}{x^+_\spm}-x^-_\spm- \frac{1}{x^-_\spm}\big) \mp 2\,q \log \frac{x^+_\spm}{x^-_\spm}=\frac{2i}{\hh} \ , \\
\end{equation}
The variables $x'^{\pm}_+$ and $x'^{\pm}_-$ are simply given by sending $p\to
p'$ and $\e_\pm \to \e_\pm'$. Solving for $x^{\pm}_+$ and $x^{\pm}_-$ in terms
of $p$ we find
\begin{equation}\begin{split}
x^{\pm}_+&\,=\frac{e^{\pm i \frac{p}{2}}(1+ q\,p + \e_+(p))}{2\, \hh \sqrt{1-q^2}\, \sin\frac{p}{2}} \ , \qquad \quad
x^{\pm}_-=\frac{e^{\pm i \frac{p}{2}}(1- q\,p + \e_-(p))}{2\, \hh \sqrt{1-q^2}\, \sin\frac{p}{2}} \ ,
\\ & \qquad \qquad \e_\pm =\sqrt{(1\pm q\,\hh\, p)^2+4\,\hh^2(1-q^2)\sin^2\frac{p}{2}}\ . \label{xqpm}
\end{split}\end{equation}
As expected, at leading order in the near-BMN expansion the dispersion relation
is given by $e_\pm$ as defined in \eqref{eq:dr}. The functions $\eta_\pm$ and
$\nu_\pm$ are generalized in the obvious way from \eqref{etanu}.

In section \ref{sec:phq} we are interested in expanding the functions $x^\pm_+$
and $x^\pm_-$ at strong coupling. To do so it is convenient to introduce new
variables $x_\pm$ such that
\begin{equation}\begin{split}\label{xqexp}
x^\pm_+= & \, x_+ \pm \frac{i}{h} \frac{x_+^2}{\sqrt{1-q^2}(x_+^2-1)- 2\, q\, \hh\, x_+}+\mathcal{O}(\hh^{-3})\ , \qquad
\\ x^\pm_- = & \, x_- \pm \frac{i}{h} \frac{x_-^2}{\sqrt{1-q^2}(x_-^2-1)+ 2\, q\, \hh\, x_-}+\mathcal{O}(\hh^{-3})\ .
\end{split}\end{equation}
Expressing $x_\spm$ in terms of $p$ in the near-BMN expansion (i.e. first
rescaling $p$) one finds
\begin{equation}\label{xqofp}
x_\spm(p)=\frac{1 \pm q \,p+\sqrt{(1\pm q\,p)^2+(1-q^2)p^2}}{\sqrt{1-q^2}\,p}+\mathcal{O}(h^{-2})\ .
\end{equation}

\section{Phase factors for \txpf{$AdS_3\times S^3\times M^4$}{AdS3 x S3 x M4} backgrounds}\label{phases}

In this appendix we give the relevant details regarding the dressing phases for
the RR backgrounds and their expansion in the near-BMN limit.

In eqs.~\eqref{funpp} and \eqref{funpm} $S^{11}_{++}(p,p')$ and
$S^{11}_{++}(p,p')$ appeared as overall phase factors in the exact result. In
the case of $AdS_3\times S^3\times T^4$ their expressions are known exactly
\cite{Borsato:2013hoa}
\begin{align}
S^{11}_{++}(p,p')^{-1}&=e^{-\frac{i}{2} a (\e' p-\e p')}\sqrt{\frac{x'^- - x^+}{x'^+ - x^-}\, \frac{1-\frac{1}{x^+ x'^-}}{1-\frac{1}{x^- x'^+}}} \frac{\n'}{\n} \, e^{i\, \vartheta^{11}_{++}(x^\pm,x'^\pm)}\ ,\label{Spp}\\
S^{11}_{+-}(p,p')^{-1}&=e^{-\frac{i}{2} a (\e' p-\e p')}\sqrt{\frac{1-\frac{1}{x^+ x'^+}}{1-\frac{1}{x^- x'^-}} \frac{1-\frac{1}{x^+ x'^-}}{1-\frac{1}{x^- x'^+}}}\, \n' \, e^{i\, \vartheta^{11}_{+-}(x^\pm,x'^\pm)}\ .\label{Spm}
\end{align}
The functions $\vartheta^{11}_{++}(p,p')$ and $\vartheta^{11}_{+-}(p,p')$ can
be expressed in terms of an auxiliary function $\chi$
\begin{align}
\vartheta^{11}_{++}(x^\pm,x'^\pm)&=\chi(x^+,x'^+)+\chi(x^-,x'^-)-\chi(x^+,x'^-)-\chi(x^-,x'^+)\ ,\\
\vartheta^{11}_{+-}(x^\pm,x'^\pm)&=\tilde\chi(x^+,x'^+)+\tilde\chi(x^-,x'^-)-\tilde\chi(x^+,x'^-)-\tilde\chi(x^-,x'^+)\ ,
\end{align}
and the explicit all-order expressions for $\chi$ and $\tilde \chi$ are
\begin{align}
\chi(x,y)&=\chi^{\BES}(x,y)+\frac12\big(-\chi^{\HL}(x,y)+\chi^-(x,y)\big)\ ,\\
\tilde\chi(x,y)&=\chi^{\BES}(x,y)+\frac12\big(-\chi^{\HL}(x,y)-\chi^-(x,y)\big)\ .
\end{align}
Here the function $\chi^{\BES}$ is the same as that which appears in the
$AdS_5\times S^5$ dressing factor \cite{Beisert:2006ez}, $\chi^{\HL}$ is the
Hernandez Lopez phase \cite{Hernandez:2006tk} and is given by the one-loop term
in the strong coupling expansion of $\chi^{\BES}$, while the function $\chi^-$
does not appear in the $AdS_5\times S^5$ light-cone gauge S-matrix. The three
functions can be expressed compactly as contour integrals
\begin{align}
\chi^{\BES}(x,y)&= i \ointc \frac{dw}{2 \pi i} \ointc \frac{dw'}{2 \pi i} \, \frac{1}{x-w}\frac{1}{y-w'} \log{\frac{\Gamma[1+i \hh(w+1/w-w'-1/w')]}{\Gamma[1-i \hh(w+1/w-w'-1/w')]}}\ ,
\label{eq:besdhmrep}\\
\chi^{\HL}(x,y)&= \frac{\pi}{2} \ointc \frac{dw}{2 \pi i} \ointc \frac{dw'}{2 \pi i} \, \frac{1}{x-w}\frac{1}{y-w'} \, \text{sign}(w'+1/w'-w-1/w)\ , \\
\chi^-(x,y) &=\ointc \, \frac{dw}{8\pi} \frac{1}{x-w} \log{\left[ (y-w)\left(1-\frac{1}{yw}\right)\right]} \, \text{sign}((w-1/w)/i) \ - x \leftrightarrow y \ .
\end{align}

We are interested in the near-BMN expansion of these expressions. Therefore,
let us quote the first two orders of $\vartheta^{11}_{++}(x^\pm,x'^\pm)$ and
$\vartheta^{11}_{+-}(x^\pm,x'^\pm)$
\begin{align}\label{strong}
\vartheta^{11}_{++}(x^\pm,x'^\pm)&=\frac{1}{h}\vartheta^{\AFS}(x,x')+\frac{1}{h^2}\vartheta_{++}^{(1)}(x,x')+\mathcal{O}(h^{-3})\ ,\\
\vartheta^{11}_{+-}(x^\pm,x'^\pm)&=\frac{1}{h}\vartheta^{\AFS}(x,x')+\frac{1}{h^2}\vartheta_{+-}^{(1)}(x,x')+\mathcal{O}(h^{-3})\ , \label{strong2}
\end{align}
The functions appearing in \eqref{strong} and \eqref{strong2} are given by
\begin{align}
\vartheta^{\AFS}(x,y)&=\frac{2 (x-y)}{(x^2-1)(xy-1)(y^2-1)}+\mathcal{O}(h^{-2})\ , \nonumber \\
\vartheta_{++}^{(1)}(x,y)&=\frac{1}{\pi}\frac{x^2}{x^2-1}\frac{y^2}{y^2-1}\left[\frac{(x+y)^2 (1-\frac{1}{xy})}{(x^2-1)(x-y)(y^2-1)} + \frac{2}{(x-y)^2} \log\left(\frac{x+1}{x-1}\frac{y-1}{y+1}\right)\right]+\mathcal{O}(h^{-1})\ ,\nonumber \\
\vartheta_{+-}^{(1)}(x,y)&=\frac{1}{\pi}\frac{x^2}{x^2-1}\frac{y^2}{y^2-1}\left[\frac{(xy+1)^2 (\frac{1}{x}-\frac{1}{y})}{(x^2-1)(xy-1)(y^2-1)} + \frac{2}{(xy-1)^2} \log\left(\frac{x+1}{x-1}\frac{y-1}{y+1}\right)\right]+\mathcal{O}(h^{-1})\ .\label{AFSstrong}
\end{align}
It is important to point out that the pre-factors appearing in \eqref{Spp} and
\eqref{Spm} can be written as a phase factor whose exponent has a vanishing
one-loop ($\mathcal{O}(h^{-2})$) term. This property, together with
\eqref{AFSstrong}, allows us to compare $\vartheta^{(1)}_{++}$ and
$\vartheta^{(1)}_{+-}$ directly with our perturbative result following from
unitarity methods.

The same property also holds for $AdS_3\times S^3 \times S^3\times S^1$, in
which case we have a total of four undetermined phase factors
\begin{align}
S^{\a\a}_{++}(p,p')^{-1}&=e^{-i a (\e' p-\e p')} \frac{1-\frac{1}{x^+ x'^-}}{1-\frac{1}{x^- x'^+}}\frac{x'^- - x^+}{x'^+ -x^-} \left(\frac{\n'}{\n}\right)^2 \, e^{ i\, \vartheta^{\a\a}_{++}(x^\pm,x'^\pm)}\ ,\label{SppS1} \\
S^{\a\a}_{+-}(p,p')^{-1}&=e^{-i a (\e' p-\e p')}\sqrt{\frac{1-\frac{1}{x^+ x'^+}}{1-\frac{1}{x^- x'^-}} \frac{1-\frac{1}{x^+ x'^-}}{1-\frac{1}{x^- x'^+}} }\, \n' \, e^{ i\, \vartheta^{\a\a}_{+-}(x^\pm,x'^\pm)}\ , \label{SpmS1}
\end{align}
and
\begin{align}
S^{\a\bar\a}_{++}(p,p')^{-1}&=e^{-i a (\e' p-\e p')}\ \frac{1-\frac{1}{x^+ y'^-}}{1-\frac{1}{x^- y'^+}} \frac{\n'}{\n} \, e^{ i\, \vartheta^{\a\bar\a}_{++}(x^\pm,x'^\pm)}\ ,\label{SpppS1} \\
S^{\a\bar\a}_{+-}(p,p')^{-1}&=e^{-i a (\e' p-\e p')}\ \sqrt{\frac{1-\frac{1}{x^+ y'^+}}{1-\frac{1}{x^- y'^-}}} \left(\frac{1-\frac{1}{x^+ y'^-}}{1-\frac{1}{x^- y'^+}} \right)^{\frac32}\, \n' \, e^{ i\, \vartheta^{\a\bar\a}_{+-}(x^\pm,x'^\pm)}\ . \label{SppmS1}
\end{align}
Unlike the $AdS_3\times S^3 \times T^4$ case all-order expressions for
$\vartheta^{\a\a}_{\s_M\s_N}$ and $\vartheta^{\a\bar\a}_{\s_M\s_N}$ are not
known. The one-loop near-BMN expansions for these phases are given in
eqs.~\eqref{1Lphasepp}, \eqref{1Lphasepm} and are essentially the same as
\eqref{AFSstrong} up to an overall scaling depending on the masses.

\bibliographystyle{nb}
\bibliography{AdS3}

\end{document}